\documentstyle[12pt,thesis]{report}

\input psfig

\begin{document}

\beforepreface
\prepage
\midface
\afterpreface



\chapter{Introduction}

In modern computational science, long sequences of random
numbers are required in various fields such as statistical
mechanics, particle physics, and applied mathematics. Methods
utilizing random numbers include Monte Carlo simulation
techniques \cite{Bin92}, stochastic optimization \cite{Aar89},
and cryptography \cite{Seb89}, all of which usually require
fast and reliable random number sources. In practice, the random
numbers needed for these methods are produced by deterministic
rules, implemented as pseudorandom number generators which
usually rely on simple arithmetic operations. Obviously, these
pseudorandom number sequences can be ``random'' only in some
limited sense, and therefore their main purpose is only to
{\em imitate} random behavior as well as possible. Assuming that
physical stochastic processes such as nuclear decay and thermal
noise allow us to generate ``truly'' random number number
sequences (in the sense that they do not contain correlations),
using this approach might be a more reliable method than
use of pseudorandom number sequences. However, due to practical
reasons physical sources are usually not used.

Since the very idea of using deterministic algorithms
in generation of random variables is in conflict with
any idea of randomness, an obvious question arises: how can
these sequences be used in applications such as Monte Carlo
simulations, whose performance is fully based on the assumption
of truly random numbers? In an illustrative sense, the
justification for their use may be considered in terms of
the accuracy: when the number of independent samples $N$ is
small, the precision of the Monte Carlo method
is poor (the error being proportional to $1/\sqrt{N}$ \cite{Koo90}).
Therefore, {\em subtle} deviations from randomness in pseudorandom
number sequences may not appear unless very many samples
are taken. Thus, for such computational applications
in which high precision is not a crucial requirement, there are
numerous fairly good pseudorandom number generators which will
work just fine. Such generators are like compasses for the sailors
in the $15^{th}$ century: in those days, when long distance
voyages were not a standard routine, a compass often lead the
ship close to the desired place, where other means such as
local knowledge could be utilized to find the precise location.

However, the technological development of computers has lead to a
situation, where carrying out ever demanding computational tasks
is possible. In the case of Monte Carlo simulations, this means that
in addition to studying more challenging problems, more accurate
simulations (with larger $N$) can be carried out. When such high
precision simulations are being done, however, there must be better
sources of randomness than just ``fairly'' good pseudorandom number
generators; the compass must be replaced with a satellite navigation
system. In other words, improvement of the accuracy leads
to a situation where the quality of pseudorandom number sequences
should improve as well. Otherwise, ambiguous results may appear.
For example, in the mid 1980's high precision calculations of
the critical temperature in the three-dimensional Ising model
\cite{Bax82} received a lot of attention, and in the cases where
dubious results were found, the quality of some pseudorandom
number generators was questioned \cite{Bar85,Hoo85,Hoo83,Parisi85}.
This raises another question: how can we determine that a
pseudorandom number sequence is ``random enough'' for some
particular application? Naturally, if an exact analytic answer
is known in a special case, for example, we may check the quality
of a pseudorandom number sequence {\em in situ} by using it in
this application. Otherwise, the quality of pseudorandom number
sequences must somehow be tested indirectly. Basically, there
are two such indirect means. {\em Theoretical tests}
(\cite{Knu81} pp. 75-110) are based on studying some
properties such as the period length and uniformity of a
pseudorandom number generator algorithm. However, since almost
without exception theoretical tests study the properties of
algorithms over their entire cycle, {\em empirical tests}
(\cite{Knu81} pp. 59-73) are also needed. In addition to
studying properties of finite subsequences instead of the whole
period, empirical tests may also give us further insight on how
the pseudorandom number generator behaves. Furthermore,
since the implementation of the algorithm (written for a
computer) may be incorrect, empirical tests provide the means
to detect possible errors.

In the course of time, many theoretical
\cite{Cov67,Knu81,Lec93,Nie89,Tez87} and empirical
\cite{Blu84,But61,Gar78,Goo53,Gre55,Gru51,Ken38,Ker37,Lec92,Mar85}
\cite{Mau92,Nai38,Sch93,Ugr91,Yue77,Yul38} tests
for pseudorandom number generators have been suggested. However,
since all pseudorandom number generators are based on a
deterministic algorithm, it is always possible to construct a test
for every generator where it will fail. Therefore, although the
success of pseudorandom number generators in extensive testing
improves confidence in their properties, it is never a sufficient
condition for their use in all applications. Recently, this
phenomenon was observed in some high precision Monte Carlo simulations,
in which several commonly used pseudorandom number generators
gave incorrect results \cite{Cod93,Fer92,Gra93a,Gra93b,Sel93} when
special simulation algorithms were employed. Still, these
generators have performed well in several earlier tests
\cite{Chi87,Ham94,Kir81,Mar91,Vat93}. Since it is
important to make sure that a pseudorandom number generator
is good enough for a chosen application, it is important to
test it with such tests, which mimic the properties of the
application where the generator will be used. In other words,
efficient {\em application specific tests} of
randomness are clearly needed.

The aim of this work is to present five new tests
for pseudorandom number generators. These tests have been
developed from the point of view of a physicist,
in the sense that they are based on direct analogies to some
physical systems such as the Ising model \cite{Bax82}, which
has been utilized in the first two tests. {\em The
cluster test} \cite{Kankaala93} is based on comparing the
cluster size distribution of a random lattice with the
Ising model at an infinite temperature. Then, in {\em the
autocorrelation test} \cite{Vat94a} we calculate the
integrated autocorrelation time of some quantities of the
Ising model, when the Wolff updating method \cite{Wol89a} is
being used. In addition, two other tests related to random
walks will also be proposed. In {\em the random walk test}
\cite{Vat94a}, we consider the distribution of the final position
of a random walk on a plane which is divided into four equal
blocks. {\em The $n$-block test} \cite{Vat94a} is based on the
idea of renormalizing a sequence of uniformly distributed random
numbers, and it is essentially a random walk test in one dimension.
Finally, {\em the condition number test} \cite{Vat94c} utilizes
some exact results on Gaussian distributed random matrices.

The outline of this Thesis is as follows. In Chapter 2, we first
consider the concept of randomness, which because of its
extraordinary character has no unique definition. We proceed
by considering generation methods of randomness and their desired
properties mainly from a practical point of view,
and therefore the emphasis of this discussion is on
pseudorandom number sequences. Then, a brief historical
perspective of random number generation leads us to
Chapter 3, where pseudorandom number generators and their use
are studied in more detail. Test methods for randomness
are discussed in Chapter 4, where more detailed descriptions
of the tests developed in this work are also presented.
Following this background, the results of these tests are
given in Chapter 5, where we first demonstrate that the
cluster test is particularly powerful in finding periodic
bit level correlations, being the most efficient of three
bit level tests whose efficiency we have studied in this
work. The other two bit level tests are included in this
work for the purpose of comparison only. Moreover, we show
that the autocorrelation test, the
random walk test, and the $n$-block test are very effective
in detecting short-range correlations, whereas the results
of the condition number test are mostly inconclusive.
Finally, the summary and discussion are given in Chapter 6.


\chapter{Concept of randomness}

Randomness may be regarded as a notion opposite to being deterministic,
which means that if the history of some process is known then its
future may be predicted. Therefore, in random phenomena no memory
effects should be present, and hence consequtive events should be
{\em independent} of each other \cite{ENCYCL}. Despite this clear
description, there are fundamental problems in the actual definition
of randomness. Although this problem concerns mostly mathematicians,
it has also relevance in several applications such as cryptography
and reliability of modern Monte Carlo simulations. Therefore,
in the beginning of this Chapter we will briefly discuss the
current situation concerning the definition of randomness.
An extensive discussion for infinite sequences is given by
Knuth (\cite{Knu81} pp. 142-161). For more recent reviews, see
{\em e.g.} van Lambalgen \cite{Lam87} and Compagner \cite{Com91b}.

Due to the essential requirement of random behavior, {\em i.e.}
the independence of consequtive events, random numbers should be
produced by measuring a stochastic process such as radioactive
decay or flipping a fair coin. When computer simulations are concerned,
however, this is not very practical. Therefore, for practical
purposes random numbers are generated by using completely
deterministic rules, so called pseudorandom number generators.
In order to compare these two approaches, we will consider the
pros and cons of both methods and give reasons for preferring
deterministic methods in computational applications. Furthermore,
the desired properties of random number sources will also be
considered. For good reviews on random number generation see
{\em e.g.} Jansson (\cite{Jan66} pp. 22-68), Anderson \cite{And90},
and L'Ecuyer \cite{Lec93} and for desired properties of random
number sequences see James \cite{Jam90} and L'Ecuyer \cite{Lec93},
for example. Also, due to its long and interesting history,
the main steps in the development of generation methods of
random numbers will also be considered. An interested reader
is referred to Hull and Dobell \cite{Hul62}.

\bigskip
\section{Definitions of randomness}

Despite its simple meaning to a layman, for mathematicians
the definition of {\em randomness} has caused a lot of worry
and despair. Still, in spite of all the work done, the definition
of randomness is not unique but the discussion on this subject
is still in progress. In the following, we will consider
randomness mainly in terms of two formal concepts: complexity
and unpredictability. Furthermore, since the formal approaches
are not very useful for practical purposes, we will also consider
two more practical approaches to define randomness.

For a layman, randomness is usually related to such practical
ideas as irregularity and unpredictability. Regularity may be
easily illustrated visually, like considering the distribution
of (motivated) soldiers in a marching order;
physicists may consider the spatial
distribution of spins in the two-dimensional Ising model in its
ground state at zero temperature. As an opposite notion,
the distribution of trees in a forest in its natural state might
seem irregular from the point of view of a layman.
The notion of unpredictability is also clear. For example,
if all the winning numbers of lottery are collected for
several years and are then statistically analysed, one should
not be able to predict following winning numbers better than
by flipping a fair coin.

For mathematicians, on the other hand, defining randomness is not
as simple. A major advancement for its understanding occurred in
1919, when von Mises introduced the notion of {\em Kollektiv},
standing for a single infinite sequence of random events
\cite{Com91b}. In his work he also attempted to set a foundation
for probability theory. For a recent review on von Mises' work,
see Ref. \cite{Lam87}. Later in the 1960's, Kolmogorov \cite{Com91b},
Martin-L\"of \cite{Martin66}, and Chaitin \cite{Cha66} described
random sequences in terms of complexity. In the course of time,
this discussion has led to the identification of randomness
with polynomial-time unpredictability \cite{Lec89}, a required
property {\em e.g.} in cryptography \cite{Lec89}. Later,
Compagner \cite{Com91a,Com91b} has proposed defining
randomness in the case of binary sequences in terms of
being uncorrelated.

The essence of the work of Kolmogorov, Martin-L\"of \cite{Martin66},
and Chaitin \cite{Cha66,Cha75,Cha75b,Cha88} is based on the idea
that the information embodied in a random piece of data cannot be
reduced to a more compact form. For example, consider two sequences
$\{ x_i\}$ and $\{ y_i \}$ of zeros and ones, $i=1,\ldots,30$:

\medskip
\begin{tabular}{c c c c c}
 & & $\{ x_i \}$ & = & 100100100100100100100100100100 \\
 & & $\{ y_i \}$ & = & 101101101111010101101110001000 \\
\end{tabular}
\medskip

The sequence $\{ x_i\}$ can be written in a more compact form as
``repeat 100 ten times'', whereas for the sequence $\{ y_i\}$
such ``compression'' is not possible. Hence, for the random
bit sequence $\{ y_i \}$ the shortest way to write it is to give
each element explicitly. The idea of {\em complexity}
is based on this compactivity: it is defined as the length
of the shortest program on a Turing machine (a universal
but formal binary computer \cite{Tur37}) that produces the
binary sequence \cite{Com91b}. As a result, the definition of
randomness may be written as follows \cite{Com91b}:
{\em ``A binary sequence is random if its complexity is
not smaller than its length.''} Naturally, this approach
is not restricted to binary sequences only.

In the course of time, the idea of complexity has lead
to another formal way of defining randomness. The key
notion in this approach is {\em unpredictability}, which
due to its importance in many practical applications such as
cryptology has recently received a lot of attention
\cite{Blu84,Boy89,Boy89b,Kan88,Kra92,Sch93}. This
approach is based on the ideas of computational complexity
\cite{Pap94} (for a review see {\em e.g.} L'Ecuyer \cite{Lec93}),
where we consider the time it takes to guess the next element $x_{n+1}$
of the sequence $\{ x_i \}$, $i=1,\ldots,n$, when the entire past
is known. Randomness in the sense of unpredictability is
then satisfied, if no polynomial time algorithm (in size
of the sequence) can guess the next element
significantly better than by flipping a fair coin
\cite{Lec89}. As an example, consider the following three sequences
of bits:

\medskip
\begin{tabular}{c c c c c}
 & & $\{ x_i \}$ & = & 100100100100100100100100100100 \\
 & & $\{ y_i \}$ & = & 101101101111010101101110001000 \\
 & & $\{ z_i \}$ & = & 010011010111000100001111011001 \\
\end{tabular}
\medskip

In the case of the sequence $\{ x_i \}$, it takes just a moment to say
that the $31^{st}$ element is most probably 1. For the sequence
$\{ y_i \}$ we are not able to predict the ``correct'' value, since
this sequence has been produced by asking 30 different people to give
arbitrarily one of two values: one or zero. The sequence $\{ z_i \}$
also appears ``random'' at first sight, but further investigations
may reveal that it has been generated by using a well-defined and simple
formula \cite{Ehr92}. Therefore, we may assume that $\{y_i\}$ is the
only sequence which would satisfy randomness from the point of view
of unpredictability.

Then, let us consider previous formal definitions from a practical
point of view. Let us assume that we have a random number sequence,
which must be tested against the hypothesis that it obeys (at least)
one of these definitions; {\em i.e.} it is ``random'' or it is not. In
the case of the notion of complexity, we note that since the number
of possible programs increases exponentially with its length
\cite{Wolfram85}, and each program of progressively greater length
must be tried in order to find the shortest one, and any one of
them may run for an arbitrarily long time, testing this approach is
clearly impractical. For unpredictability some tests have also
been developed \cite{Blu84,Sch93}, but besides being very tedious
tasks to perform, to our knowledge no such tests have even been
carried out. Therefore, when testing randomness is needed,
more practical approaches to define randomness must be considered.

One such approach has been studied by Compagner and coworkers
\cite{Com91a,Com91b,Com87}. They considered finite binary
sequences and proposed testing the values of all possible
correlation coefficients of an ensemble of a given sequence.
They then suggested that the essential requirement for randomness
is {\em uncorrelatedness}; {\em i.e.} the disappearance of all
the correlation coefficients \cite{Com91a}. Unfortunately,
although this approach gives a well-defined way to test
randomness, it is obvious that even this definition appears
rather formidable for practical purposes.

Another more practical definition of randomness has been provided
by Lehmer \cite{And90}, whose definition essentially describes
randomness from the empirical point of view: a random sequence is
``{\em a vague notion embodying the idea of a sequence in
which each term is unpredictable to the uninitiated and whose
digits pass a certain number of tests, traditional with
statisticians and depending somewhat on the uses to which the
sequence is to be put.}'' The underlying reason for the use of
this definition is simple: since no unique practical recipe for
testing a finite sequence of numbers has been given \cite{Lam87},
various authors have developed different tests
\cite{Blu84,But61,Cov67,Gar78,Goo53,Gre55,Gru51,Ken38,Ker37}
\cite{Lec92,Mar85,Mau92,Nai38,Sch93,Tez87,Ugr91,Yue77,Yul38}
which probe some properties of the sequences. Then, if a
random number sequence passes several well chosen tests, nothing
of its ``randomness'' is proven but the confidence towards its
properties increases. Moreover, if the chosen tests mimic the
properties of an application in which the random number sequence
will be used, this criterion may be good enough for practical
purposes.

As a brief conclusion of this Section we may note that
regardless of the application in which random numbers are
used, their quality must be tested by some means.
In this sense, although some test can be constructed
for all the aforementioned definitions, the approach
of Lehmer \cite{And90} to find support for random behavior
by conducting several {\em practical} and
well chosen tests is the most suitable one. Therefore,
from now on, we will consider randomness from the point of
view of Lehmer's definition. Instead of
discussing tests, however, we will now briefly consider
two different types of randomness which such test methods
study. In Chapter 4, we will return to testing random
number sequences and consider that subject in more detail.

\bigskip
\section{Global and local randomness}

Let us briefly consider two possible types of randomness
in a random number sequence: global and local randomness.
Since the classification between the two is more or less
obscure, in this work we try to avoid confusion by defining
them as follows. Consider a random number sequence $\{ x_i \}$,
$i=1,2,\ldots,N_T$ with $N_T \gg 1$. By means of some test
for randomness, we study its properties over $n \leq N_T$ elements
in this sequence. {\em Global} properties are studied when
$n/N_T \sim {\cal O}(1)$. On the other hand, when
$n/N_T \ll 1$, {\em local} properties are considered. This
definition follows the ideas of Kendall and Babington-Smith,
who were the first to introduce the concept of local properties
of randomness in the 1930's \cite{Ken41,Ken38}. Moreover,
since the test is not explicitly specified, this definition
is general in the sense that it is not restricted to the
Lehmer's definition for randomness \cite{And90} only.

There is one point which must be further specified: global
randomness does not guarantee realization of local randomness.
To illustrate differences between local and global properties,
let us consider uniformity in a sequence $\{1,2,,\ldots,100\}$
of 100 integers. In global sense, this sequence is uniformly
distributed since every number between one and 100 occurs
exactly once, but when uniformity of the first ten numbers are
studied, it certainly is not (in the range $1,2,\ldots,100$).
Another interesting example concerns truly random sequences,
by which we mean sequences where no correlations are present.
In such (finite) sequences local randomness is not necessarily
realized \cite{Knu81} whereas global randomness is.

\bigskip
\section{Generation methods of randomness}

Since random numbers have applications in many fields,
a variety of different generation methods have been developed,
each with their own set of advantages and disadvantages.
Depending on the application, three types of methods are
usually used: truly random, pseudorandom, and quasirandom
sequences.

{\em True randomness} corresponds to ideally random behavior,
meaning that in a truly random sequence of numbers no correlations
are present, as already mentioned in the previous Section.
Usually, one attempts to generate truly random numbers by
measuring some (physical) stochastic process such as
radioactive decay \cite{Ino83} or thermal noise \cite{Ric92}.
For that reason, these sources are sometimes also called
physical random number generators. Although this method
allows generation of truly random numbers in principle, in
practice, however, it is troubled with several
problems such as bias due to human preference for certain
digits \cite{Ken39}, human errors in measurements
\cite{Hel94,Ken38}, or ignorance of the correlation time
in the system measured \cite{War87}. Moreover, although
there are methods to improve the properties of
the output of physical random number generators {\em e.g.}
by extracting most random bits of the output \cite{Cho88,Neu63},
this method is still too slow for use in most computational
applications. Hence, in practice truly random numbers are
generated only for specific purposes such as lottery
\cite{Tho59} and testing in algorithm development
\cite{Ede_comm,Vat93}. Also, tables (see {\em e.g.} Refs.
\cite{Ken61,Que59,Sha62,Tip27} and \cite{Jan66} pp. 23-26) and
some high capacity storage devices \cite{Ric92} have been
made for further use of truly random numbers.

On modern computers, instead of using truly random numbers,
several alternative methods have been developed for generating
sequences which are {\em not} random but try to {\em imitate}
random behavior for simulation purposes. These so called
{\em pseudorandom} number sequences are produced by
deterministic algorithms, implemented as pseudorandom number
generators which usually rely on simple arithmetic
operations. Despite the obvious conceptual conflict concerning
deterministic algorithms in production of random number
sequences, in many computational applications in which
pseudorandom numbers are used, real problems seldom occur
{\em provided that} some care is taken to ensure that the
pseudorandom number generator is ``good'';  {\em i.e.} it
has passed several well chosen tests. When such ``good''
pseudorandom number generators are used, many computational
applications in which their output is utilized are fairly
robust for subtle (but inevitable) deviations from randomness.
The problems usually appear only when high
precision simulations and some special algorithms
\cite{Cod93,Fer92,Gra93a,Gra93b,Mos93,Sel93,Zif92_pre} are used.
However, a word of warning must be given here. In the course of
time, hundreds (or even thousands) of pseudorandom number
generators have been suggested, but the theoretical and
empirical properties are {\em well} known only for few. As a matter
of fact, generators known to be bad are certainly still used in
several computing centers around the world.
As James \cite{Jam90} has pointed out, there
are ``{\em many internal reports devoted to the revelation that
the local `official random number generator' is not random
enough}''. Hence, since pseudorandom number sequences are
not truly random, most results based on the use of these
sequences must be taken with a sceptical attitude.

Roughly speaking, {\em quasirandom} number sequences are defined
as sequences of points, whose purpose is not to even imitate
true randomness, but to estimate a given problem with
as small an error as possible. A good example is numerical Monte
Carlo integration \cite{Koo90}. There, one can estimate the
integral for some particular function $f$ by taking a sample
of $N$ points over the integration domain. Then, the average of
$f$ at those points, multiplied by the volume of the integration
domain, gives an estimator for the integral. Now, if the points
are taken from a truly random or pseudorandom sequence of numbers,
the error will be proportional to $1/\sqrt{N}$ \cite{Koo90}.
But one can do better if the sample points are spread
throughout the integration domain ``more evenly'' \cite{Lec93}
than in the case of truly random sequences. Such quasirandom
number sequences are tailored to satisfy equidistribution criteria
better than truly random and pseudorandom number sequences,
resulting in an error proportional to $1/N$ \cite{Kle92}.
However, despite the obvious importance of quasirandom
sequences in some applications such as interpolation problems
and the numerical solution of integral equations \cite{Nie78},
in this work we will not consider them any further. For a review
of quasirandom number methods see, {\em e.g.}, Niederreiter
\cite{Nie78}. One practical implementation is given in
Ref. \cite{Bra88}.

For the purpose of completeness, let us also consider two other
possible sources of randomness: transcendental numbers and
chaos. For a long time decimals of {\em transcendental numbers} such
as $\pi$ and $e$ have been considered random, and for that
reason they have been calculated in various occasions
\cite{Gru52,Rei50,Sha62}. In several studies, no excessive
deviations from randomness have been
found\footnote{\protect\samepage{It has been argued
(\protect\cite{Jan66} p. 38)
that the decimals of $e$ mimic truly random behavior too well,
since in some tests the empirical distribution
has been found to follow the theoretical one too closely.
However, due to a small number of decimals studied no conclusions
can be drawn.}} \cite{Gru52,Met50,Pat62}, but
they still share the same problem with arithmetic algorithms:
they are deterministic. In this sense these numbers are also
pseudorandom numbers. Moreover, since the cost for the
computation of the decimals of $\pi$, for example, gets
progressively higher when more decimals are needed, this
method is clearly impractical. {\em Chaos}, on the other
hand, can be effectively utilized as a source of randomness.
This idea is meaningful, since chaos denotes a state of
disorder and irregularity \cite{Schuster84}, which are common features
of randomness as well. In practice, this connection has been
made use of by developing pseudorandom number generators
\cite{Lus94,Pha93} which are based on the theory of deterministic
chaos \cite{Schuster84}. In spite of its general interest, in
this work we will not consider this idea any further.

In the remaining part of this work, only the terms of truly
random and pseudorandom number sequences will be used. Moreover,
when it is not essential to specify which sequence we are
considering, we will speak of random number sequences
in general.

As we have noticed, all three methods discussed here have their
own place in the sea of applications. Truly random number
sequences are needed as long as people want to gamble.
Pseudorandom number sequences are useful in simulations
where properties like speed and repeatability are essential.
These and other desired properties of random number sequences
in computational applications are discussed in more detail in
the following Section. Finally, in applications in which uniformity
is the most essential requirement, quasirandom number sequences
are very useful. Therefore, let us quote Knuth \cite{Knu81}:
``{\em We are forced to conclude that no sequence of `random'
numbers can be adequate for every application}.''

\bigskip
\section{Desired properties of random number sequences}

There are several properties which are required for or at least
desired from random number generators used in modern computational
applications \cite{Jam90,Lec93}. Since a computational approach
of physical problems is our main interest, we will
consider these properties in some detail. Furthermore, for
the purpose of completeness, we will also consider realization of
these properties in the case of physical sources of random numbers.

Naturally, the most important property is ``{\em good randomness}''.
In the sense of Lehmer's definition, a pseudorandom number sequence
should then pass a well chosen set of tests before extensive use
in some particular application. As we mentioned above, physical
random number generators do not necessarily produce truly random
sequences, and therefore their output must be tested as well. The final
criterion of goodness of a random number sequence, however, is
determined by the application: if the random number sequence
does not give the correct answer within error limits, it is
not random enough. Problems arise, if no such check against
analytic results is possible. Then, a simple check of
randomness is to calculate one typical problem with
few different random number generators, and compare
their results.

Another often desired property of random number sequences is
{\em a uniform distribution}. Although this is not essential, it
is of considerable importance since most nonuniform distributions
can be formed by using uniformly distributed random numbers
between zero and one \cite{Bra83}. In this work, only uniform
random number generators will be considered. For an introductory
survey of computer generation of nonuniform distributions, see
Ripley \cite{Rip83}. A library of {\sc Fortran} routines for this
purpose is given in Ref. \cite{Bro94}.

For physical random number sequences, {\em a long period} is not a
problem due to their aperiodic nature. For pseudorandom number
sequences, however, this may be a problem since almost all pseudorandom
number sequences are finite and reproducible. Aperiodic algorithms
have also been suggested \cite{Ehr92,Ziel90}, but they are troubled
either with poor results in empirical tests \cite{Ehr92} or with
computational difficulties \cite{Ziel90}. In the case of periodic
pseudorandom number sequences, there are occurrencies in which
the existing long-range correlations may be avoided by using only
a small portion of the whole period \cite{Mat90}. Furthermore,
in some applications such as parallel simulations, the output
of the generator is split into numerous disjoint subsequences,
which should be independent of each other. These facts emphasize
the need for extremely long periods, so that only a small
portion of the entire cycle needs be used in a single
simulation. The longer the period, the better.

Also, depending on the application, the efficiency of random
number generation may be important. Most pseudorandom number
generators currently used in modern computational applications
(in supercomputer environments) produce about $10^6$ or more
random numbers per second, and therefore in most cases the
question of efficiency may be neglected. Its importance arises
mainly in high precision Monte Carlo simulations in which huge
amounts of pseudorandom numbers are used. In the case of physical
sources of random numbers this question is not insignificant either.
For example, in 1983 one such method \cite{Ino83} was able to
generate about 600 physical random numbers in a second, when
the storing time to a magnetic tape was included. Clearly this
is too slow for use in real time.
Of course, the time can be shortened if the machine is used
on-line, but then the results of the calculation would no
longer be repeatable. In computer simulations, however,
{\em repeatability} is a crucial requirement mostly because of
testing purposes: sometimes it may be necessary to repeat
the simulation with exactly the same random number sequence.
For pseudorandom number generators this requirement
is usually fulfilled.

Finally, from the point of view of computer simulations,
portability and parallelizability are often required.
In general, {\em portability} means that when a random
number sequence has been generated on some particular machine,
there must be means to generate exactly the same sequence
on other machines also. For pseudorandom number
generators this is usually not a problem, if high level
programming languages (like {\sc Fortran} or {\sc C}) and good
programming techniques are used. In the case of physical random
numbers, however, this property is fulfilled only if they are
first stored and then transferred. When huge numbers of
random numbers are needed, this certainly becomes impractical.
{\em Parallelizability} has become important with the
development of parallel computers. In practice, parallelizability
means that pseudorandom number sequences generated on different
processors are independent of each other. For certain
classes of pseudorandom number generators this is possible
\cite{Edd90}, although problems with independence have also
appeared \cite{Mos93}. In the case of physical sources of random
numbers, consideration of parallelizability does not
make much sense.

\bigskip
\section{Brief history of random number generation}

Random number generation and testing is a very widely studied
subject. Prior to 1979 hundreds of articles had been published
already (for an extensive biography, see Refs.
\cite{Nance72,Sah79,Sow72,Sow78}), and ever since their
number has been continuously increasing. A rapid increase
of articles related to this subject occurred in the 1950's,
hand in hand with the development of computers which needed
practical and reliable sources of random numbers. However,
before describing any further the development of arithmetic
methods used in computers, we will consider the main steps
leading up to the computer era.

Means for generating ``randomness'' have been known for a long
time. For example, many games rely on the random nature of well
shuffled cards and dice, when the possibility of cheating is
excluded, of course. Roulette is another famous example, but
unfortunately empirical results of Pearson \cite{Pea00} are
against its random nature: in 1900, he analysed a series of
runs of colour (red or black) in the throws of the roulette
ball in Monte Carlo, and concluded that the odds are at least
$10^{30}$ to 14.5 against such a series! Other methods have also
been tried. In the 1920's, Tippett \cite{Tip27} collected about
40\,000 random digits of census reports. Amazingly, although
these digits were tested with several tests, no evidence for
nonrandomness were found \cite{Dod42,Ken38,Nai38,Yul38}.
Later, Kendall and Babington-Smith \cite{Ken38} attempted to
construct a random sequence by selecting digits from the London
Telephone Directory. However, it was found that the series
was significantly biased, and therefore the London Telephone
Directory was ``{\em useless as a source of random digits}''
\cite{Ken38}. On the other hand, Kermack and McKendrick \cite{Ker37}
also tested certain telephone numbers which were found favourable
to randomness. This result was commented by Kendall and
Babington-Smith that ``{\em Kermack and McKendrick are
apparently dealing with a five-figure Scottish exchange}''
\cite{Ken38}.

Collecting data of census reports or throwing a dice
is a slow process, however. Therefore, mechanized machines
were built to produce numbers more efficiently.
The first such work was done by Kendall and Babington-Smith
\cite{Ken38,Ken39}, who generated a table of 100\,000 random
digits by using a rapidly spinning disk divided into ten
equal sections. This work is a remarkable token of patience,
since all the 100\,000 digits were collected by Babington-Smith
only, with a rate of 1500 digits per hour on the average.
Moreover, in four tests only about 5\% of the digits
were considered suspicious. Probably the most famous
example of physical random number generation is the book of
one million random digits and 100\,000 normal deviates,
published by the RAND Corporation in 1955 (see reviews in
Refs. \cite{Mul55,Tom56}). These digits were generated by a kind
of automatic ``electronic roulette'', and they have passed
several tests after minor adjustments \cite{Jan66}.
Later, special machines for purposes
of lottery have been introduced. A classical example is ERNIE
(Electronic Random Number Indicator Equipment) \cite{Tho59},
which was used by the British General Post Office
to pick the winners in the Premium Savings Bond Lottery.
Naturally, the output of ERNIE was tested with several
tests, which revealed no excessive
deviations from randomness \cite{Tho59}.  We are not
aware if this machine is still in operation.

Recently, to our knowledge, only few studies of physical
random number generators have been published. In the 1980's,
Inoue {\em et al.} \cite{Ino83} described a method for
generating random digits by measuring decay from
radioactive nuclei. Richter \cite{Ric92} generated a
vast amount of random digits by means of measuring thermal
noise from a semiconductor device. These digits have been
permanently stored on a computer disk, from which they can
be transferred when needed.

Because of practical reasons, however, physical methods were
quickly replaced by arithmetic methods as soon as
the development of computers took place. Hence, the history
of random number generation during the computational
era is closely related to the development of computers.

Generally speaking, the Monte Carlo method denotes any method
in which random numbers are used \cite{Jam80}. Hence it is very clear
that the history of these two fields of science, pseudorandom
number generation and Monte Carlo methods, are closely related
to each other. As a matter of fact, since the development of the
first electronic computer, ENIAC \cite{Gol46}, many of the people
who took part in its development and use, have had a strong
influence in these fields of science.
For example, N. Metropolis originally suggested
\cite{Met87} an obvious name for the Monte Carlo simulation method
\cite{Met49,Met53}, and with J. von Neumann studied randomness of the
decimals of $\pi$ and $e$ \cite{Met50} and developed the
first algorithm for generating pseudorandom numbers: the so
called midsquare method \cite{Hul62,Knu81}. In this method,
an arbitrary $n$-digit integer is squared, creating a
$2n$-digit product. A new integer is formed by extracting
the middle $n$ digits from the product. Although the
properties of random numbers generated with the midsquare method
are bad \cite{Knu81} (consider 50 with $n=2$, for example),
it was still used in the 1960's \cite{Fos63}.

At the same time, another arithmetic method was suggested
by Lehmer: the so called multiplicative linear congruential
generator \cite{Leh51}. This method was already used on ENIAC,
and due to its simplicity and well understood theoretical
properties it is still widely used. Ever since,
development of computers and Monte Carlo algorithms
have set continuously increasing demands for the properties
of pseudorandom number generators. As a result, several
new classes of pseudorandom number generators have been
suggested during the last few decades. These generators
and their properties are the subject of next Chapter.


\chapter{Pseudorandom number generators}

In this Chapter, we will consider the properties and use of
pseudorandom number generators. First, most commonly used
pseudorandom number generator algorithms will be presented,
the emphasis being on their most important factors such as
their structure and known properties. In this context,
we will not pay attention to the particular values of the
parameters in these algorithms but consider their
properties in general. In addition,
we shall also consider few other promising methods, which
may prove useful in the near future. Reviews of current
state of generation methods can be found {\em e.g.} in
Anderson \cite{And90}, James \cite{Jam90} (see also
corrections in Refs. \cite{Jam94,Lus94}), and L'Ecuyer
\cite{Lec90,Lec93}.

In the next Section, we shall describe in more detail the
generators (with specified values for the parameters), which
have been chosen for the tests developed in this work
(cf. Chapter 4). In order to better serve the physics
community, we have tried to choose those
algorithms which are widely used or otherwise seem
promising, and which have been previously tested.
Thus, we are able to summarize the current understanding
of their properties, and since developing new,
more accurate tests is one of our main objectives, we
are then able to compare our new tests with tests
developed by other authors.

Finally, even the best pseudorandom number generator algorithm
can be defeated by an incorrect computer implementation, improper
use, or bad initialization. These and other practical questions
shall be studied in the last Section of this Chapter, where
some tips for avoiding them shall also be given.

\bigskip
\section{Classification of pseudorandom number generators}

Most commonly used pseudorandom number generator algorithms are
the {\em linear congruential method}, the {\em lagged Fibonacci method},
the {\em shift register method}, and {\em combination methods}.
Other methods such as the {\em add-with-carry} and
{\em subtract-with-borrow} generators and improvements of
previous methods have also been proposed. In the following,
main features of these methods will be given.

\bigskip\bigskip\noindent
{\bf Linear congruential generators}
\medskip

\noindent
Among the simplest algorithms are the {\em linear
congruential generators} which use the integer recursion
\begin{equation}
X_{i+1} = (a\ X_{i}\ +\ c\ )\mbox{ mod }m,
\end{equation}
in which the integers $a$, $c$ and $m$ are constants. It generates
a sequence $X_1, X_2, \ldots$ of random integers between
0 and $m-1$ (or in the case $c = 0$, between 1 and $m-1$).
Each $X_{i}$ is then scaled into the interval [0,1).
If the multiplier $a$ is a primitive root modulo $m$ (and
$X_0 \neq 0$ in the case $c=0$) and $m$ is prime, the period
of this generator is $m-1$. For other cases, the period length is
given in Ref. \cite{And90}, but then the low order bits are not
random. Linear congruential generators can be classified into
{\em mixed} $(c > 0)$ and {\em multiplicitive} $(c = 0)$
types, and are usually denoted by LCG$(a,c,m)$ and MLCG$(a,m$),
respectively.

Since the introduction of this algorithm by Lehmer \cite{Leh51},
its properties have been studied in detail. Marsaglia \cite{Mar68}
pointed out about 25 years ago that the random numbers in $d$
dimensions lie on a relatively small number of parallel hyperplanes.
This lattice structure was further studied by Ripley \cite{Rip83b}.
Boyar (see Ref. \cite{Boy89} and references therein)
proved that LCG generators are efficiently (in polynomial time)
predictable when the constants $a$, $c$, and $m$ are unknown.
Despite these deficiencies, in simulations LCG generators are
widely used, and therefore a vast amount of theoretical
work \cite{Cov67,Fis90,Fis82,Fis86,Mat92} has been done to weed
out bad choices of these constants.

\pagebreak

\noindent
{\bf Fibonacci method}
\medskip

\noindent
To increase the period of the linear congruential algorithm,
it is natural to generalize it to the form
\begin{equation}
X_{i} = (a_{1}\ X_{i-1}\ +\cdots +\ a_{r}\ X_{i-r})\mbox{ mod }m,
\end{equation}
in which $r > 1$ and $a_{r} \neq 0$.
The period is the smallest positive integer $\lambda$ for which
\begin{equation}
(X_{0},\ \ldots ,\ X_{r-1}) = (X_{\lambda},\ \ldots ,\ X_{\lambda+r-1}).
\end{equation}
Since there are $m^{r}$ possible $r$-tuples, the maximum
period is $m^{r} - 1$. The use of
$r\ =\ 2,\ a_{1}\ = \ a_{2}\ = 1$ leads to
{\em the Fibonacci generator}
\begin{equation}
X_{i} = (X_{i-1}\ +\ X_{i-2})\mbox{ mod }m.
\end{equation}
Since no multiplications are involved, this implementation has the
advantage of being fast. Due to its poor properties,
however, the Fibonacci generator is rarely used (\cite{Knu81} p. 26).

\bigskip\bigskip\noindent
{\bf Lagged Fibonacci generators}
\medskip

\noindent
A natural extension to the Fibonacci method is the {\em lagged
Fibonacci} generator, which requires an initial set of elements
$X_{1}, X_{2}, \ldots , X_{r}$ and then uses the integer recursion
\begin{equation}
X_{i} = (X_{i-r}\ \otimes\ X_{i-s})\mbox{ mod } m,
\end{equation}
in which $r$ and $s$ are two integer lags satisfying $r > s$
and $\otimes$ is one of the binary operations
$\{+,-,\times,\oplus\}$, $\oplus$ being an exclusive-or
operation. The corresponding generators are designated by
LF($r,s,m,\otimes$). Usually the binary operation is addition
or subtraction modulo $2^{w}$, $w$ being the word length (in
bits). Then, the maximal period with suitable choices of $r$
and $s$ is $(2^r - 1)2^{w-1} \approx 2^{r+w-1}$ \cite{Mar85}. When
multiplication (with odd integers) or exclusive-or are used,
the period lengths are $(2^r - 1)2^{w-3} \approx 2^{r+w-3}$
and $2^r - 1$, respectively \cite{Mar85}.

Excluding knowledge of the period length, theoretical properties
of lagged Fibonacci generators (in terms of $r$ and $s$) are
not deeply understood, which makes their recommendation quite
difficult. An exception are the lagged Fibonacci generators
based on the exclusive-or operation, but they will be considered
in the context of the generalized feedback shift register generators.
For a relative goodness of the operations $\{+,-,\times,\oplus\}$,
some empirical results are fortunately known. First, according to
Marsaglia \cite{Mar85}, exclusive-or should never be used.
Results of Coddington's high precision Monte Carlo simulations
\cite{Cod93} then suggest that multiplication is the best choice
of the operations $\{+,-,\times,\oplus\}$, although it gives
a shorter maximal period than addition and subtraction.
For further details of LF generators see, {\em e.g.},
Refs. \cite{Mar85,Mar85b}.


\bigskip\bigskip\noindent
{\bf Tausworthe generators}
\medskip

\noindent
An alternative generator type is the {\em shift register generator}.
Feedback shift register generators are also sometimes called
Tausworthe generators \cite{Tau65}, which are based on the
theory of primitive trinomials of the form $x^{p} + x^{q} + 1$
\cite{Gol82}. Given such a primitive trinomial and $p$ binary
digits $x_{0}, x_{1}, x_{2},\ldots , x_{p-1}$, a binary shift
register sequence can be generated by the following recurrence
relation:
\begin{equation}
x_{i} = x_{i-p}\ \oplus\ x_{i-q},
\end{equation}
in which $\oplus$ is the exclusive-or operator, which is equivalent to
addition modulo 2. $b$-bit words can be formed from bits taken
from this binary sequence as
\begin{equation}
W_{j} = x_{jb}\ x_{1+jb} \cdots x_{(b-1)+jb},
\end{equation}
in which $j=0,1,2,\ldots$.
The resulting binary words are then treated as random numbers.
Such a sequence of random integers will have the maximum
possible period of $2^{p} - 1$, if $x^{p} + x^{q} + 1$ is a
primitive trinomial and if this trinomial divides $x^{n} - 1$
for $n = 2^{p} - 1$, but for no smaller $n$. These conditions
can be met by choosing $p$ to be a Mersenne prime, {\em i.e.}
a prime number $p$ for which $2^{p} - 1$ is also a prime.
A list of Mersenne primes can be found {\em e.g.} in Refs.
\cite{Her92,Kur91,Zie69,Zie68}. Also, Tezuka \cite{Tez90} has shown
that Tausworthe sequences form structures similar to lattice
structure of linear congruential sequences \cite{Mar68}.
Finally, let us just mention that results of some empirical
tests suggest that generators based on small
values of $p$ should not be used \cite{Mar85}, and the value
of $q$ should be small or close to $p/2$ \cite{Too71}.

\pagebreak

\noindent
{\bf Generalized feedback shift register generators}
\medskip

\noindent
Generalization of Tausworthe sequences has been suggested
by Lewis and Payne \cite{Lew73}. They formed $b$-bit words
by introducing a delay between the words. The corresponding
generator is called the {\em generalized feedback shift register}
generator, denoted by GFSR$(p,q, \oplus)$. In a GFSR generator
with two lags $p$ and $q$ the words $W_{i}$ satisfy
the recurrence relation
\begin{equation}
\label{Eq:GFSR}
W_{i} = W_{i-p} \oplus W_{i-q}, \mbox{\hspace{1.0cm}} p > q,
\end{equation}
which clearly shows that the GFSR generator is a special case
of lagged Fibonacci methods. Hence, with properly chosen
lags  \cite{Her92,Kur91,Zie69,Zie68} maximal period length
of $2^{p}-1$ can be achieved for GFSR generators.

An important aspect of the GFSR algorithm concerns its
initialization, in which $p$ initial seeds are required: in the
least fortunate case, if the $j^{th}$ bit is zero in each
of the first $p$ integers of the sequence, it will remain zero
throughout. Theoretically this question has been studied in
Refs. \cite{Fus83b,Fus88,Fus89,Fus83a,Tez87,Tez87b}.
Based on theoretical studies
\cite{Andre90,Arv78,Fus90,Gol82,Lew73,Nie87,Tez87,Tez88},
GFSR generators have good properties in general, although
the lattice structure observed for Tausworthe sequences is also
a problem for GFSR generators \cite{Tez90}. Golomb (\cite{Gol82}
pp. 78-79) has theoretically shown that the decimation
of a maximum-length GFSR sequence by powers of two\footnote{By such
decimation we mean that only every $k^{th}$ number ($k$ being a power
of two) of the sequence produced by Eq. (\ref{Eq:GFSR}) is used.}
results in equivalent sequences.
Moreover, based on Ref. \cite{Com87} the correlation
length $\xi$ of GFSR generators equals the lag parameter
$p$. This results from the so called three-point (triple)
correlations by which we mean (trivial) correlations of the form
$x_{i}\ \oplus\ x_{i-p}\ \oplus\ x_{i-q} = 0$, in which
$x_{i}$'s denote single bits in the words $W_{i}$.
Such correlations dominate the properties of GFSR
generators (over their full period), as Ziff has shown
in an unpublished work \cite{Zif92_pre}.

\bigskip\bigskip\noindent
{\bf Decimation of GFSR sequences and primitive pentanomials}
\medskip

\noindent
As mentioned above, the values of lags $p$ and $q$ determine the
properties of GFSR generators (with two lags). In order to
improve their properties there are two possibilities: one may
increase the values of lags $p$ and $q$, or one may increase
their number. In an unpublished work \cite{Zif92_pre}, Ziff
considered the latter possibility, developing GFSR generators
with four lags:
\begin{equation}
\label{Eq:Decimation}
W_{i} = W_{i-p} \oplus W_{i-q_1} \oplus W_{i-q_2} \oplus W_{i-q_3},
\end{equation}
in which $p > \max (q_1,q_2,q_3)$. Such a generator will be denoted by
GFSR($p$,$q_1$,$q_2$,$q_3$). The theory underlying the choice of
lags $p$, $q_1$, $q_2$, and $q_3$ is given in Ref. \cite{Zif92_pre},
and is based on the decimation of Eq. (\ref{Eq:GFSR}) with
some value of $k$ which is not a power of two (such as $k=3,5,7$).
For that reason, this method is also called {\em $k$-decimation} of
GFSR$(p,q,\oplus)$. Moreover, although the parameter $p$ in
Eqs. (\ref{Eq:GFSR}) and (\ref{Eq:Decimation}) is the same, the values
of lags $q_1$, $q_2$, and $q_3$ are determined from the value of $k$.

The period of the $k$-decimated sequence is also $2^p - 1$.
In addition to theoretical studies of Ziff \cite{Zif92_pre},
his empirical studies \cite{Zif92_pre} indicate that this method
really improves the properties of GFSR generators.

Despite the decimation of GFSR generators with two lags, these
decimated sequences are still based on primitive trinomials.
Another way to increase the number of lags is to develop
generators based on {\em primitive pentanomials}
of the form $x^{p} + x^{q_1} + x^{q_2} + x^{q_3} + 1$.
Some values for the lags $p > q_1 > q_2 > q_3 > 0$ are proposed in
Ref. \cite{Kur91}, which allows construction of an another set
of generators like Eq. (\ref{Eq:Decimation}). To our knowledge
no studies for such generators have been carried out.

\bigskip\bigskip\noindent
{\bf Combination methods}
\medskip

\noindent
Given the inevitable dependencies that will exist in a pseudorandom
number sequence, it seems natural that one should try to {\em shuffle}
a sequence \cite{Ito92} or to {\em combine} separate sequences.
An example of such approach is given by MacLaren and Marsaglia
\cite{Mac65} who were apparently the first to suggest the idea
of combining two generators together to produce a single sequence
of random numbers. The essential idea is that if
$X_{1},\ X_{2},\ldots $ and $Y_{1},\ Y_{2},\ \ldots $ are
two random number sequences, then the sequence $Z_{1},\ Z_{2},\ \ldots$
defined by $Z_{i}\ =\ X_{i}\ \otimes\ Y_{i}$ will not only be ``more
uniform than either of the two sequences but will also be ``more
independent'' \cite{Mar85}. The symbol $\otimes$ mentioned above is
one of the binary operations $\{+,-,\times,\oplus\}$. Algorithms
using this idea are often called {\em mixed} or {\em combination}
generators.

Despite strong empirical support for combination methods
\cite{Alt88,Den91a,Ham94,Lec88,Lec92,Mar85,Mar68b,Mar90,Mar94,Tez91,Wes67},
their theoretical understanding is still limited.
Usually, the period of the combination is much longer than
that of its single components \cite{Lec88,Mar90,Mar85}, but
only few theoretical studies suggest that combination really
improves properties such as uniformity and ``independence''
\cite{Brown79,Mar85,Den91b}. However, as L'Ecuyer has
pointed out \cite{Lec93}, ``statistical defects'' are a common
problem of many fast and simple generators such as LCG and
GFSR generators, which when combined could yield an efficient
generator with much better properties. Further
theoretical studies of the combination method can be found
in Refs. \cite{Lec91,Tez94,Tez91}.

\bigskip\bigskip\noindent
{\bf Other methods}
\medskip

\noindent
Finally, let us consider few other generation methods of
randomness, which may prove useful in the near future.
The generators proposed by Marsaglia and Zaman \cite{Mar91}
will be considered in some detail, whereas some other methods
shall only be mentioned. Moreover, since these methods will be
considered only here, theoretical and empirical test
results for some particular generators will also be given.

Recently, Marsaglia and Zaman \cite{Mar91}
have proposed the so called {\em add-with-carry} (AWC) and
{\em subtract-with-borrow} (SWB) generators. These generators
are basically lagged Fibonacci generators, with an extra
addition of the carry bit (AWC) or subtraction of the borrow
bit (SWB). Their main advantage is a very long period, the smallest
for the generators suggested in Ref. \cite{Mar91} being
approximately $10^{171}$. Fairly soon, however, defects in
these generators were found. Tezuka {\em et al.} \cite{Tez93}
proved these generators to be equivalent to LCG's with
large moduli, and therefore they have an unfavourable lattice
structure \cite{Cou94,Tez93}.
In addition, some results of SWB generators in empirical tests
\cite{Lec92,Vat93} and high precision Monte Carlo simulations
\cite{Fer92} do not support their general use in their current
form. Recently, some light to this problem has been given by
L\"uscher \cite{Lus94}, who has suggested a way to improve
the properties of one particular SWB generator called RCARRY
\cite{Jam90} by neglecting some of the generated random numbers.
An actual implementation of this improved generator has been
given by James \cite{Jam94}.

Recently, Marsaglia has continued his previous studies and
proposed a so called {\em multiply-with-carry} (MWC)
generator \cite{Mar94MWC}. Although knowledge of its properties
is still incomplete, Marsaglia states that \cite{Mar94MWC}
``{\em all bits of the integers produced by this new method,
whether leading or trailing, have passed extensive tests of
randomness}.'' The so called {\em inversive congruential generators}
have also received considerable attention. For a review of their
properties see Ref. \cite{Eic92}. For cryptologic purposes
some nonlinear generators have also been proposed, the most
known being the so called {\em BBS generator} \cite{Blu86}. This
work has been extended {\em e.g.} in Refs. \cite{Bea94,Mic91,Reif88}.
Finally, Matsumoto and Kurita \cite{Matsu92} have proposed a
variant of GFSR generators, known as {\em twisted GFSR}.

\bigskip
\section{Tested pseudorandom number generators}

In this Section, we shall describe in more detail the generators
which have been chosen for the tests. We will focus on the main
details, properties, and drawbacks of these generators, including
some discussion on open problems. In the end, we will present
a short summary of their relative goodness.

\bigskip\bigskip\noindent
{\bf GGL}
\medskip

\noindent
GGL is a uniform random number generator based on the linear
congruential method \cite{Par88}. The form of the generator
is MLCG($16807,2^{31}-1$) or
\begin{equation}
X_{i+1} = (16807\ X_{i}) \mbox{ mod } (2^{31} - 1),
\end{equation}
and it generates pseudorandom numbers between 1 and $2^{31}-2$
(initial seed value $X_0 = 0$ is forbidden).
This generator has been particularly popular \cite{Par88}.
It has seen extensive use in the IBM computers \cite{IBM71},
and is also available in some commercial software packages
such as subroutine RNUN in the IMSL library \cite{IMSL89}
and subroutine RAND in the MATLAB software \cite{MATLAB91}.

MLCG($16807,2^{31}-1$) generators are quite fast and have
been argued to have ``highly satisfactory'' properties \cite{Lew69}.
The main disadvantage of MLCG($16807,2^{31}-1$) is its poor
lattice structure in low dimensions ($d=2,3$) \cite{And90,Lec88},
which explains very poor results in some recently
developed empirical tests \cite{Lec92}. In other
empirical tests these generators have performed well
\cite{Gar78,Kir81,Lea73,Lew69,Ugr91}, and results of several bit level
tests also support its good properties \cite{Alt88,Vat93}.
Another drawback of MLCG($16807,2^{31}-1$) is its period
$2^{31} - 2$ ($\approx 2 \times 10^{9}$ steps) \cite{Kir81},
which can be exhausted fast on a modern high speed computer.

\pagebreak

\noindent
{\bf RAND}
\medskip

\noindent
RAND uses the linear congruential method with
a period of $2^{32}$ \cite{convex} to return successive
pseudorandom numbers in the range from 0 to $2^{31} - 1$.
The generator is LCG($69069,1,2^{32}$) or
\begin{equation}
\label{rand}
X_{i+1} = (69069\ X_{i}\ +\ 1) \mbox{ mod } 2^{32},
\end{equation}
and our implementation of this algorithm is equivalent
to the implementation by Convex Corp. on the Convex C3840
computer system \cite{convex}, in which the sign bit is
always set equal to zero.

The multiplier 69069 has been used in many generators, probably
because it was strongly recommended in 1972 by Marsaglia
\cite{Mar72}, and is part of the famous SUPER-DUPER combination
generator \cite{And90}. Known properties of LCG($69069,1,2^{32}$)
do not support its use, however. Although its test results
have been fairly good \cite{Lea73,Mar85}, in higher dimensions
($d\geq 6$) its lattice structure is poor \cite{And90}, and
only its most significant bits have passed bit level tests \cite{Vat93}.
The last property is due to the modulus, which is a power of two:
the least significant bit has a period of two, the
second least significant a period of four, and so on
\cite{Edd90}. In addition, because of the poor bit level
properties of LCG($69069,1,2^{32}$)
both most and least significant bits of SUPER-DUPER
are also correlated \cite{Alt88}.

\bigskip\bigskip\noindent
{\bf RAN3}
\medskip

\noindent
RAN3 follows the algorithm of a lagged Fibonacci generator
LF(55,24,$m$,$-$) or
\begin{equation}
X_{i}  = (X_{i-55} - X_{i-24}) \mbox{ mod } m.
\end{equation}
This algorithm was originally Knuth's suggestion \cite{Knu81}
for a portable routine but with an add operation instead of
a subtraction. This was translated to a {\sc Fortran}
implementation by Press {\it et al.} \cite{Pre89} who
chose $m = 10^9$ for RAN3. The period length of RAN3 is
$2^{55} - 1$ \cite{Knu81}, and it requires an initializing
sequence of $55$ numbers. Based on results of some empirical
tests \cite{Vat93}, RAN3 has fairly good properties,
although both its most and least significant bits are
correlated \cite{Vat93}. Furthermore, in recent simulations
of three-dimensional self-avoiding random walks a
LF(55,24,$m$,$+$) gave incorrect results \cite{Gra93b}.

\bigskip\bigskip\noindent
{\bf R{\em p}}
\medskip

\noindent
In this work, generalized feedback shift register generators
GFSR($p$,$q$,$\oplus$) \cite{Lew73} with two lags $p$ and
$q$ will be denoted by R{\em p}. The value of $q$ shall be
given explicitly when necessary. These generators follow an
algorithm given by Eq. (\ref{Eq:GFSR}), and suggested values
for lags are given {\em e.g.} in Refs.
\cite{Her92,Kur91,Zie69,Zie68}.

One particular example of GFSR generators is R250 \cite{Kir81},
which generates 31-bit integers through a recurrence of
the form GFSR(250,\-103,\-$\oplus$) or
\begin{equation}
X_{i} = X_{i-250}\ \oplus\ X_{i-103}.
\end{equation}
Our implementation of this algorithm is done by Helin
\cite{Hel85}, and it needs $p=250$ words of memory to store the
$250$ latest random numbers. A new term of the sequence can be
generated by a simple bitwise exclusive-or ($\oplus$) operation.
The period of R250 is $2^{250} - 1$ \cite{Kir81},
and based on several empirical tests its
properties are good \cite{Chi87,Ham94,Kir81,Vat93}.

In recent high precision simulations, however, several GFSR
generators including R250 have produced incorrect results when
special simulation algorithms have been employed
\cite{Cod93,Fer92,Gra93a,Gra93b,Sel93,Zif92_pre}.
It was suggested \cite{Fer92} that the most significant bits of
R250 are correlated, but based on a recent study \cite{Vat93}
at least the individual bits\footnote{By individual bits we mean
bits from a particular sequence $i$ in a sequence of
$b$-bit random words ($i$ being one of $1,2,\ldots,b$).}
of R250 pass many empirical tests
on bit level. Grassberger has proposed triple correlations
with a correlation length of ``$\approx 10^2 - 10^3$'' \cite{Gra93a} or
``$\approx 400$'' \cite{Gra93b}, but even his studies have not been
able to determine the correlation length precisely. Therefore,
although we have good reason to assume that the underlying reason
for poor performance of the GFSR generators in these high precision
simulations is due to the three-point correlations with a
correlation length $\xi = p$ (cf. Section 3.1),
efficient test methods for confirming this are still
lacking.

In this work, initialization of
GFSR generators was performed with 32-bit integers produced
by GGL. Other initialization methods including the one in
Ref. \cite{Fus89} were also checked, but the results (given in
Chapter 5) were unaffected.

\pagebreak

\noindent
{\bf ZIFF{\em p} \ and PENTA{\em p}}
\medskip

\noindent
In order to study the effect of $k$-decimation of
GFSR$(p,q,\oplus)$ generators, we have implemented \cite{Sta94} the
algorithm GFSR$(p,q_1,q_2,q_3,\oplus)$ \cite{Zif92_pre} or
\begin{equation}
\label{Eq:Ziffg}
X_{i} = X_{i-p} \oplus X_{i-q_1} \oplus X_{i-q_2} \oplus X_{i-q_3},
\end{equation}
whose period is $2^{p} - 1$, $p > \max (q_1,q_2,q_3)$.
Such a generator will de denoted by ZIFF{\em p}, and values of
lags $q_1$, $q_2$, and $q_3$ \cite{Zif92_pre} shall be given
explicitly when necessary. One particular generator of
this kind has been given in Ref. \cite{Zif92}. Excluding the
results of Ziff \cite{Zif92_pre}, no test results of
these generators have been available.

In addition, generators based on pritimive pentanomials have also
been constructed. These generators also follow Eq. (\ref{Eq:Ziffg})
but with a different choice of lags \cite{Kur91}. Such
generators will be denoted by PENTA{\em p}, where values of
other lags will again be given when necessary. The period of
this generator is also $2^{p} - 1$.

The initialization of these generators was performed bit by bit
by using GGL: all $b \times p$ bits in the initial seed vector
were initialized by using the most significant bits of
integers produced by GGL.

\bigskip\bigskip\noindent
{\bf RANMAR}
\medskip

\noindent
RANMAR is a combination of two different generators \cite{Jam90,Mar90}.
The first is a lagged Fibonacci generator
\begin{equation}
X_{i}  = \left\{
	\begin{array}{ll}
        X_{i-97} - X_{i-33}, & \mbox{if $X_{i-97} \geq X_{i-33};$} \\
        X_{i-97} - X_{i-33} + 1, & \mbox{otherwise,} \\
        \end{array}
	\right .
\end{equation}
\noindent
in which only 24 most significant bits are used for single precision reals.
The second part of the generator is a simple arithmetic sequence for the
prime modulus $2^{24} - 3 = 16777213$. This sequence is defined as
\begin{equation}
Y_i = \left\{
	\begin{array}{ll}
        Y_i - e,   & \mbox{if $Y_i\geq e$;} \\
        Y_i - e + f, & \mbox{otherwise,} \\
        \end{array}
        \right .
\end{equation}
in which $e = 7654321/16777216$ and $f= 16777213/16777216$.
The final random number $Z_i$ is then produced by combining
$X_i$ and $Y_i$ as
\begin{equation}
Z_{i}  = \left\{
	\begin{array}{ll}
        X_{i} - Y_{i}, & \mbox{if $X_{i} \geq Y_{i};$} \\
        X_{i} - Y_{i} + 1, & \mbox{otherwise.} \\
	\end{array}
	\right .
\end{equation}
The total period of RANMAR is about $2^{144}$ \cite{Mar90}.
A scalar version of the algorithm has been tested on bit level
with good results \cite{Mar90}. We used the implementation by
James \cite{Jam90} which is available in the Computer Physics
Communications (CPC) software library, and has been recommended
for a universal generator. This version has passed several
empirical tests \cite{Ham94,Vat93}, showing no apparent drawbacks.

\bigskip

We may conclude this Section by saying that, based on current
knowledge of the tested pseudorandom number generators,
RANMAR seems to have best properties in general.
Despite its poor lattice structure in low dimensions,
also GGL seems to have good properties, and it has
been recommended as a ``minimal standard'' generator \cite{Par88}.
R{\em p} generators have succeeded in several tests,
but due to recent incorrect results in some high precision Monte
Carlo simulations they cannot be recommended for general use.
ZIFF{\em p} and PENTA{\em p} generators may bring some light to this
problem. Finally, RAN3 and RAND seem to be the worst of the
tested generators, showing correlations both on bit level and
in other tests.

\bigskip
\section{Pitfalls in the use of pseudorandom number generators}

For several classes of pseudorandom number generators, theoretical
knowledge is already so extensive that some generators within these
classes can be recommended, if their known limitations are taken
into account. For example, due to their apparent lattice structure
linear congruential generators should not be used in lattice
simulations without great care. However, good theoretical
properties lose their significance, if incorrect practical
procedures such as incorrect implementation of the algorithms
are performed. In this Section, we will consider
few such cases.

Starting from the pseudorandom number generator algorithm,
the first mistake one can do is to {\em implement} it incorrectly
for the computer. Practically this means that the output of the
implementation (pseudorandom number generator) is not identical
with the output of the underlying algorithm; {\em i.e.} the
implementation is not exact. In addition to human error,
reasons for such behavior may include many machine dependent features
such as finite precision of real numbers, limited word size of
the computer, and numerical accuracy of mathematical functions,
but all these can be circumvented if care is taken.
Furthermore, it would be desirable that the implemented routine
were as fast as possible and performed correctly in many different
environments, {\em i.e.} it would be {\em portable}. For further
details see Refs. \cite{Gen90,Sez92} and references therein.

One example of an incorrect {\sc Fortran} implementation is
given in the Table below.
\begin{table}[htb]
\small\centering\tt
\begin{tabular}{l l l l l l}
 & REAL FUNCTION GGL(DS)           & & &   REAL FUNCTION GGL(DS) \\
 & DOUBLE PRECISION DS, D1, D2     & & &   REAL*4 DS, D1, D2 \\
 & DATA D1/2147483648.D0/          & & &   DATA D1/2147483648./ \\
 & DATA D2/2147483647.D0/	   & & &   DATA D2/2147483647./ \\
 & DS = DMOD(16807.D0*DS,D2)       & & &   DS = AMOD(16807.*DS,D2) \\
 & GGL = DS/D1			   & & &   GGL = DS/D1 \\
 & RETURN			   & & &   RETURN \\
 & END				   & & &   END \\
\end{tabular}
\end{table}
The generator on the left corresponds to a correct 32-bit
implementation of GGL, which was introduced in Section 3.2.
In this implementation, {\tt DOUBLE PRECISION} reals are
used to carry the state of the generator, and therefore the
modulo operation is also performed in {\tt DOUBLE PRECISION}.
In the second implementation on the right, the generator has
been speeded at the expense of accuracy by using {\tt REAL*4}
definitions, which improves the speed by a factor of 1.30 on
DEC 3000 AXP. A ``slight'' drawback is a change in the period
length, which decreases by a factor of about
$3.4 \times 10^{7}$: in the incorrect
implementation it is only 64.

This example emphasizes the sensitivity of the implementation
procedure. Therefore, great care is needed both in the
implementation of pseudorandom number generator algorithms and
in the use of implemented algorithms on other machines. In the
latter case, realization of portability must be checked.

A subject which is also closely related to machine
dependent features concerns {\em optimization} during the compiling
process. Although most compilers are usually reliable,
full optimization should not be used blindly, however.
A quick check of the output of the generator with several
optimization levels is very wise. Moreover, since parallel
computing has its own problems in pseudorandom
number generation (related to independence of disjoint
subsequences), great care in these matters is needed.

All pseudorandom number generators must be initialized before
their use. For linear congruential generators
{\em initialization} is not a problem, since the only restriction
is set by the limits of their operation (possible values of
output). For example, to initialize GGL, zero should not be used
as then the sequence
will remain zero throughout. For generators with a larger
recurrence length, more care must be taken to ensure that the seed
values in the initial seed vector are independent of each other.
GFSR generators have been argued to be particularly sensitive
to initialization, and therefore special initialization methods
have been proposed, a nice example being in Ref. \cite{Fus89}.
In most cases, however, simpler methods are used. One practical
method is to use a simple generator such as a good LCG to
generate the initial seed vector. Marsaglia \cite{Mar85} has
recommended constructing the seeds bit by bit using the least
significant bit of LF(3,1,32707,$-$). Computer-specific features
such as system date and time could also be used in initialization,
but despite its usefulness this procedure has no theoretical
support. Finally, using tables of truly random numbers might
be a good idea, if very long seed vectors need not be used.

The effects of poor initialization have been studied by Altman
\cite{Alt88} and Vattulainen {\em et al.} \cite{Vat93}. Altman
observed that lagged Fibonacci generators are very sensitive to
bit level correlations in the initializing sequence.
Bit level properties of GFSR generators were studied
by Vattulainen {\em et al.}, who found that when
a GFSR generator was initialized with a correlated sequence,
the correlations did not vanish by ``warming up'' the pseudorandom
number generator but seemed to persist instead.

Unfortunately, in most applications in which pseudorandom number
generators are used, their ``random'' behavior is taken for
granted. Otherwise, it is hard to understand why most of these
studies do not report the pseudorandom number generator algorithm
used in the calculations. From the point of view of comparing
results of separate studies, however, this information would be
very valuable. Therefore, in any publication where results of
Monte Carlo experiments are given, mentioning the used
pseudorandom number generator is highly recommended.

Finally, let us mention the biggest mistake in use of pseudorandom
number generators: a ``random'' choice of the generator. Before any
generator is given considerable attention for possible use,
one should have both theoretical and empirical support for
their properties.
Good theoretical properties form the starting point for
empirical studies, which must also be performed to confirm
that the implementation of this particular algorithm is
correct, and to give more confidence for its properties.
Unless these two criteria are satisfied, that particular
generator should be avoided.


\chapter{Testing randomness}

As was pointed out in Section 2.1, there must be some means
to determine the ``goodness'' of random numbers. Traditionally
this problem has been approached by probing properties of
random number sequences by means of tests for randomness.
Unfortunately, this practical approach is troubled with
several problems, the most important being that no single
(practical) test can verify realization of randomness in a
random number sequence. For that reason numerous tests probing
different manifestations of nonrandomness have been suggested,
each having its own characteristic features. When combined
together, such a more complete test program may give a better
insight on how good the properties of the tested random
number sequence really are. In this Chapter, we will first
consider the main categories of test methods for randomness:
empirical and theoretical tests, in addition to their own
subclasses. Good reviews have been given {\em e.g.}
by Knuth (\cite{Knu81} pp. 38-110) and L'Ecuyer \cite{Lec92}.

Then, in the framework of this classification scheme we will
present the main steps in the development of tests until now.
Since the number of tests suggested by other authors is already
numerous, we are forced not to pay much attention to the details
but to give their main features instead. The only exceptions
are the chi-square test, the Kolmogorov-Smirnov test, the
$d$-tuple test, and the rank test, which will later be made
use of.

The main topic of this Chapter concerns the new test methods,
which have been developed in this work. Detailed descriptions
of these tests will be given in Section 4.4, followed
by presentation of few transformation methods for normally
(Gaussian) distributed random variables which are needed
in one of the new tests.

\bigskip
\section{Classification of test methods}

Recent practice in the classification of tests for randomness
seems confusing, since a well-established and unique division
between various test methods seems to be lacking. Instead of that,
several schemes have been proposed. The classification between
empirical and theoretical tests is a welcomed exception, since
they are easily distinguished from each other. These classes
are studied in more detail below. The notions of their subclasses,
however, seem to vary from author to the other. The term
``standard tests'' is often used to mean the tests proposed in
the book by Knuth (\cite{Knu81} pp. 59-73), followed by notions
of {\em e.g.} some ``more stringent tests'' \cite{Mar85} and
``universal tests''\cite{Mau92,Sch93}. Notions of ``visual tests''
\cite{Com87,Lew73,Vat93} and ``tests on bit level''
\cite{Alt88,Mar85,Mar85b,Vat93} are also widely used. Moreover,
Vattulainen {\em et al.} have also added this confusion by
introducing the term ``physical testing'' \cite{Vat93}, denoting
tests which are based on direct analogies to physical systems such
as the Ising model \cite{Bax82} in statistical mechanics.
In this work, we will try to avoid such confusion by neglecting
the concepts ``standard'', ``more stringent'', ``universal'', and
``physical''. Instead, we will use an important notion:
{\em application specific testing}. By this we mean tests,
which mimic the most important properties of the application
in which the random number sequence will be used. The advantage
of conducting such tests is that they will yield the most
relevant information of the properties of the random number
sequence from the point of view of this particular application.
In addition, the concepts of visual and bit level tests will be
used throughout this work. The classification scheme following
this approach is schematically shown in Fig. \ref{Fig:CLASS},
which we will consider in more detail in the following.
\begin{figure}[htb]
	\vspace{5.0cm}
\caption[A schematic illustration of the classification scheme
        of tests for randomness discussed in this work.]
	{The schematic illustration of the classification scheme
	of tests for randomness discussed in this work. The set
	of all the tests is denoted by a circle, whose separate
	parts denote different subclasses.\label{Fig:CLASS}}
\end{figure}

The first step in our classification system is the division
between studies of random bits and random words, random words
consisting of several bits each. Since both random bits and
random words may be interpreted as integers, they will be
called random numbers in general. Physical sources
and some algorithms such as the Tausworthe method generate
random bits, whereas most pseudorandom number generators produce
random words. The main reason for this classification are
some applications such as combinatorics \cite{Mar85} and
cryptography \cite{Mau92}, in which good properties
of individual random bits are required. In most applications,
however, random words are usually used.

The most traditional distinction between test methods for
randomness has been done between empirical and theoretical
tests, which can usually be designed to study properties
of random bits as well as random words. In general,
{\em empirical tests} are designed to study any manifestations
of nonrandomness such as regular patterns in visual images or
deviations from the desired distribution. Since they probe the
output of the generation method regardless of its nature
(physical or algorithmic) and not the generation method itself,
empirical tests are complementary to {\em theoretical tests},
which study characteristics of the arithmetic method itself
by using number theoretic methods. Common characteristics for
theoretical tests are {\em e.g.} the period length,
distribution, and autocorrelation properties. Another difference
between these two test methods is the type of randomness they
study. Theoretical tests are global in the sense that almost
without exception they study pseudorandom number generator
algorithms over their entire cycle. Empirical tests, on the
other hand, can be designed to study both local and global
properties (cf. Section 2.2) of random number sequences.
Furthermore, while empirical tests may study any random number
sequence regardless of its source, theoretical tests are further
restricted in the sense that for each class of pseudorandom
number generator algorithms (such as LCG and GFSR generators)
their own specific set of theoretical tests must be constructed.

In principle, there is no compelling reason for classifying
the test methods any further. For our purposes, however, two
relevant concepts will be further discussed: application
specific and visual tests. Since neither of them concerns
theoretical tests in particular, they will be discussed
from the point of view of empirical tests only.

The key concept in further classification is the application
in which the random number sequence will be used. Therefore,
let us study the sequence based on the assumption that the
application is already known. By means of this approach,
we may introduce the concept of {\em application specific
testing}, whose main feature is that such tests mimic the
essential features of the application, sometimes even being
identical to it. For example, as the exact solution of
the two-dimensional Ising model is known \cite{Bax82,Ferdi69,Ons44},
for simulations where this model is studied, it can be used as a
test for randomness as well. In most cases, however, exact results
are not known. Then, one alternative approach is to measure some
relevant quantities of the application with several different
random number generators, and compare their results with each other.
Generators whose results clearly deviate from the general trend
should then be avoided, especially if results of some generators
known to be good follow this trend. Another alternative is
to construct new tests in such a way that only the most important
properties of the application are included. An example of the
latter are numerical integrations in which uniform distribution
(in the desired dimension) is the most important property of
random number sequences, and therefore an extensive uniformity
test would serve as an application specific test for this purpose.
Then, from the point of view of application specific testing,
all remaining empirical tests (which are not application specific
given a particular application) probe properties that are less
significant for this particular application. Using the previous
example, for numerical integration bit level studies of the least
significant bits would belong to this category.

The importance of application
specific testing cannot be overestimated: since the number of
empirical tests that can be conceived is practically unlimited
and still only few tests are usually applied to any sequence,
the chosen tests should include many which mimic the
crucial properties of the application as well as possible. Therefore,
effective application specific tests are clearly needed to
reveal subtle correlations in random number sequences. This
is emphasized by recent observations in some high accuracy
Monte Carlo simulations in which biased results with some
pseudorandom number generators have been found
\cite{Cod93,Fer92,Gra93a,Gra93b,Sel93}.

Finally, let us consider one more category of tests.
{\em Visual tests} are used to test spatial correlations
between random numbers by visual means. For example, when
a sequence of random bits is put on a two-dimensional
lattice, short range correlations
are easily observed. Therefore, besides being very useful
for illustrative purposes, visual tests offer a possibility
to develop more quantitative tests through interpretation
of the visualized configurations as representations of physical
systems, such as the Ising model. This has been made use
of in the cluster test, which will be explained in Section 4.5.
Furthermore, it is clear that in some cases visual tests may
correspond to application specific testing also.

\bigskip
\section{Brief review of previous work}

In this Section, we will review the development of theoretical
and empirical tests until now. Our purpose is not to give a
thorough summary but to concentrate on giving a general idea of
their characteristic features, the emphasis being on empirical
testing. Since the number of empirical tests is vast, we
will first consider them from a historical point of view,
proceeding from the beginning of the $20^{th}$ century to the
1960's. Then, more recent developments will be discussed,
followed by discussion on recent application specific tests.

\bigskip\bigskip\noindent
{\bf Theoretical tests}
\medskip

\noindent
As we noticed in Section 3.1, there are several commonly
used classes of pseudorandom number generators, each with
their own set of parameters. Since these generators are
based on deterministic algorithms, there must be means to
analyse their behavior theoretically. For some classes of
generators such theoretical tests have been designed
\cite{Cov67,Knu81,Lec93,Nie89,Tez87}.

The purpose of all these tests is to analyse the goodness
of an algorithm in terms of its parameters, based on some
chosen measure. Several measures such as discrepancy
\cite{Lec93}, serial correlations \cite{Knu81}, lattice
structure \cite{Mar68}, Walsh functions \cite{Tez87} and
period length \cite{Knu81} have been suggested.
In the following, we will concentrate on
one particular and illustrative example of theoretical
testing: the so called spectral test. For more details on
this subject see Knuth (\cite{Knu81} pp. 75-110) and
L'Ecuyer \cite{Lec93}.

The main drawback of LCG generators was first observed in
1967 by Coveyou and MacPherson \cite{Cov67}. They found that
LCG generators exhibit a lattice structure in the sense
that the points generated by LCG generators lie on a set
of equidistant parallel hyperplanes. One example of a such
structure is shown on the left in Fig. \ref{Fig:plots},
in which 5\,000 2-tuples (consequtive pairs ($x_{2i},x_{2i+1}$)
of random numbers, $i=0,1,2,\ldots$) of GGL are plotted.
\begin{figure}[htb]
	\vspace{5.0cm}
\caption[2-tuples of GGL and GFSR(17,3,$\oplus$).]
	{On the left we have 5\,000 2-tuples of GGL, and on the
	right 1000 2-tuples generated by GFSR(17,3,$\oplus$).
	The horizontal part has been expanded for illustration
	purposes. Lattice structure in both cases is clearly
	observed. \label{Fig:plots}}
\end{figure}
Coveyou and MacPherson also gave an algorithm for computing the
distance $l_d$ between such parallel hyperplanes in $d \geq 2$
dimensions and called it {\em the spectral test}. In this test,
the shorter the distance $l_d$ between the hyperplanes the
better the generator. Later these calculations were performed more
explicitly by Marsaglia \cite{Mar68}, and an algorithm for the
computation of $l_d$ was proposed by Knuth (\cite{Knu81} pp. 98-101).
For higher dimensions ($d>10$), a more efficient algorithm has
been proposed by Fincke and Pohst \cite{Fin85}.

When good LCG generators are being searched for, the number of
possible combinations of parameters $a$, $b$, and $m$ is
practically unlimited. For that reason theoretical support for
their choice given by the spectral test is invaluable, and
therefore it has been used on many occasions \cite{Fis90,Fis82,Fis86}
to weed out bad choices. Since similar tests for
other classes of generators have not been constructed,
the spectral test has remained the most significant single
achievement in theoretical testing of pseudorandom number
generator algorithms.

Recently, similar lattice structures have been observed for other
classes of generators also. The AWC and SWB generators described
in Section 3.1 are equivalent to LCG generators, and therefore
also have structural properties \cite{Cou94} that can be analysed
by studying the lattice structure of their LCG representations
\cite{Tez93}. Furthermore, as Tezuka \cite{Tez90} has pointed out,
Tausworthe and GFSR generators also exhibit certain lattice
structure, as we may notice on the right in Fig. \ref{Fig:plots}.

\bigskip\bigskip\noindent
{\bf Empirical tests}
\medskip

\noindent
Apparently the first empirical tests for randomness were
the chi-square ($\chi^2$) and the Kolmogorov-Smirnov (KS) tests,
which were introduced by Pearson in 1900 \cite{Pea00} and by
Kolmogorov originally in 1933 \cite{Knu81}, in respective order.
Strictly speaking, these two tests are {\em methods} rather than
tests, since very often they are not used as such but utilized
by other empirical tests. In this work, however, we will follow
the standard practice and call them tests. Since they
are also used in connection with the tests developed in this
work, a more detailed description of the chi-square and KS tests
will be given in Section 4.3.

Other well-known tests for randomness were
proposed in 1938 by Kendall and Babington-Smith \cite{Ken38}.
These tests for random digits were the frequency test, the
serial test, the poker test, and the gap test. The frequency
test is based on the idea that all the numbers in a random
number sequence should occur an approximately equal number
of times. In the serial test we count the number of times
that the pairs of successive random numbers ($X_i$,$X_{i+1}$)
occur, and then calculate their deviation from the uniform
distribution. A correction to the proposed version of the
serial test has later been given by Good \cite{Goo53}.
These two tests are still widely
used, and should be included in any reasonable test bench.
The poker test is based on an analogue of the card game
having the same name, and the gap test examines the length
of ``gaps'' between occurrencies of $X_i$ in a certain
range. Later Gruenberger \cite{Gru50} described a computational
approach for some of the tests proposed by Kendall and
Babington-Smith.

At about the same time there appeared several other works related
to empirical testing. Kermack and McKendrick \cite{Ker37,Ker37b}
proposed the ``run'' test, in which segments of successive digits
with increasing or decreasing length are examined. Despite some
errors in the original proposition, the run test has later been
corrected and used extensively \cite{Knu81}. Nair \cite{Nai38}
described a test based on some ideas of Pearson \cite{Pea33}, and
Yule \cite{Yul38} studied normally distributed random variables
which were formed by adding five uniformly distributed random
variables together. The next flood of new empirical tests
followed the development of ENIAC. In 1951 the $d^2$ test was
proposed by Gruenberger and Mark \cite{Gru51}, in 1955 the coupon
collector's test by Greenwood \cite{Gre55}, and in 1961
the partition test by Butcher \cite{But61}. Many of these
and several other tests are described in more detail by
Knuth (\cite{Knu81} pp. 38-73), and practical implementations
of several such tests have been given by Dudewitz and Ralley
\cite{Dud81}.

Development of pseudorandom number generators has lead
to a situation where many modern generators pass most of
the aforementioned tests. Therefore, new tests
have been designed to detect the deficiencies of many poor
generators. This development occurred mostly in the 1980's.
A well-known example of this development is Marsaglia's
\cite{Mar85} test bench DIEHARD, which contains a set of
eight new tests. These tests have been applied in several
studies with great success \cite{Alt88,Lec92,Mar85,Vat93}, in
the sense that many generators known to pass other tests
have failed them. In addition, L'Ecuyer \cite{Lec92} has
described the nearest pair test, which especially LCG
generators tend to fail. For cryptologic purposes Blum and
Micali \cite{Blu84} have proposed the next-bit-test, which
measures the ability to predict the next bit from the
preceeding ones. This work was extended by Schrift and Shamir
\cite{Sch93}. Other tests have been proposed
{\em e.g.} by Yuen \cite{Yue77}, Garpman and Randrup
\cite{Gar78}, Ugrin-\v{S}parak \cite{Ugr91}, and Maurer
\cite{Mau92}.

Although several of the tests mentioned above can be performed
to test random bits as well, specific tests for this purpose
have also been proposed. In DIEHARD, two tests are well suited
for this purpose. In the rank test \cite{Mar85b}, samples of
the rank of a random binary matrix are taken, and their
probability distribution function is then compared with the
expected behavior. The $d$-tuple test (also known as the
overlapping $m$-tuple test) \cite{Alt88,Mar85,Vat93} is
basically a run test performed on strings of bits in random
numbers. These tests are desribed in more detail in
Section 4.4.

A rather different way of testing spatial correlations between
random numbers is possible by using direct visualization. This can
most easily be done in two dimensions by plotting pairs of random
numbers on a plane like in Fig. \ref{Fig:plots}, or visualizing
the bits of binary numbers \cite{Com87,Vat93}, as in
Fig. \ref{Fig:BITrand}.
\begin{figure}[htb]
\vspace{5.0cm}
\caption[Bits of random numbers generated by RAND.]
	{Bits of random numbers generated by RAND, as put on the
	lattice of size $50\times 50$ one bit after another. Black
	squares denote ones and white squares zeros. Starting from
	the left, the figures correspond to the $27^{th}$, $24^{th}$,
	and $21^{st}$ bits of successive random numbers, bit number
	one denoting the most significant bit. \label{Fig:BITrand}}
\end{figure}
In Fig. \ref{Fig:BITrand}, the development of irregularity
of individual bits is easily seen. Other methods
for visual testing have been given {\em e.g.} in
Refs. \cite{Green85,Lew73,Rob82,Vat93}.

\bigskip\bigskip\noindent
{\bf Application specific tests}
\medskip

\noindent
The development of application specific testing started more
or less in the 1980's, when several papers on physical simulation
models as tests for randomness appeared. Above all, the Ising
model has been the most popular in this respect. First, Kirkpatrick
and Stoll \cite{Kir81} studied pseudorandom number generators by means
of the two-dimensional Ising model with the standard Metropolis
scheme, followed by other groups \cite{Bar85,Hoo85,Hoo83,Kal84,Parisi85}
who utilized the three-dimensional Ising model. Other Monte Carlo
checks of the quality of pseudorandom number generators include
the $\phi^4$ model \cite{Mil86}, the two-dimensional Ising model
with the Wolff algorithm \cite{Cod93,Fer92,Sel93}, and non-biased
\cite{Bin88,Zif92_pre} and self-avoiding \cite{Gra93a,Gra93b}
random walks. In addition to these physical models, Lewis and Payne
\cite{Lew73} have proposed a simple ``scattering simulation test'',
Landauer \cite{Lan84} has considered queueing systems,
and Paulsen \cite{Pau84} has performed Monte Carlo time
series simulations. Moreover, some of the tests in DIEHARD
\cite{Mar85} such as the rank test and the parking lot test
have also a definite application, their results
having use {\em e.g.} in graph theory \cite{Mar85}
and lattice simulations, respectively.

Although this brief list is certainly far from complete, it
gives certain idea of the approach for application specific testing.
Popularity of the Ising model in these tests is not surprising
due to its extraordinary status in statistical mechanics.
As a matter of fact, two of the new tests proposed in
Section 4.5 are also very closely related to the Ising model.

\bigskip
\section{Chi-square and Kolmogorov-Smirnov tests}

In the following, brief descriptions of the chi-square test and
the Kolmogorov-Smirnov test will be given. For further details
see Refs. \cite{Knu81} pp. 39-56, \cite{Rub81}, pp. 26-30, and
\cite{Vat93}. Although these two tests can be used to study any
random sample, in this work we will consider them from the
computational point of view. Therefore, we assume that we
have some random number generator, whose output we are
about to study.

\bigskip\bigskip\noindent
{\bf Chi-square test}
\medskip

\noindent
One of the most common tests for randomness is {\em the chi-square
test}, which is widely used in connection with other tests.
In general, it is performed in the following way.

Let us take $N$ independent observations
$X_1,X_2,\ldots,X_N$ from some distribution generated by a
random number generator, and assume that every observation $X_i$
can fall into one of $\mu$ mutually exclusive categories
with probability $p_i$, $i=1,2,\ldots,\mu$. Then let
$O_i$ be the number of observations that actually do fall
into the category $i$. Also, let us form the test statistic
\begin{equation}
\label{Eq:chi}
V = \sum_{i=1}^{\mu} \frac{(O_i - N p_i)^2}{N p_i}
\end{equation}
with $\nu = \mu - 1$ degrees of freedom. Then consider a null
hypothesis $H_0$ that the generator is good if $V$ obeys the $\chi^2$
distribution, and consider the {\em observed} descriptive level
$\alpha = \mbox{Prob}( \chi^2 \leq V \, | \, H_0)$, $\alpha \in [0,1)$.
When $V$ (and thus $\alpha$) is very small, the empirical
distribution follows the theoretical one too smoothly in the
statistical sense, and the null hypothesis $H_0$ must be
rejected: the random number generator fails the test. On the
other hand, if $V$ (and $\alpha$) is very large, the empirical
distribution is too far from the theoretical one and again
the generator fails. Choice of the failing criteria is up to
the statistician, but usual choices are domains below
$\alpha = 0.05$ and above $\alpha = 0.95$. For the purpose
of illustration, in Fig. \ref{Fig:Chi} we show the (cumulative)
$\chi^2$ distribution with $\nu$ degrees of freedom with four
percentage points $p$ and over a range of $\nu$.
\begin{figure}[htb]
\vspace{5.0cm}
\caption{The $\chi^2$ distribution with $\nu$ degrees of freedom
	with four percentage points $p$ and over a range of $\nu$.
	The acceptance and rejection regions of the two-sided
	chi-square tests at levels of significance 0.02 (solid
	lines) and 0.10 (dotted lines) are also shown.
	\label{Fig:Chi}}
\end{figure}

When the chi-square test is applied, the statistician has to
decide the number of independent observations $N$. Parameter
$N$ must be large enough, since the statistic $V$ in Eq. (\ref{Eq:chi})
obeys the $\chi^2$ distribution {\em approximately}, the
approximation being the better the higher the values of
products $N p_i$ is. As a rule of thumb, Knuth
\cite{Knu81} recommends $N p_i \geq 5$ for all $i$.

\bigskip\bigskip\noindent
{\bf Kolmogorov-Smirnov test}
\medskip

\noindent

The disadvantage of the chi-square test is that it requires
binning of data. This drawback is eliminated in the
{\em the Kolmogorov-Smirnov} (KS) {\em test}, in which such
binning is not needed.

In the KS test, we make $N$ independent observations of some
random quantity $X$ and compare their empirical cumulative
distribution function (cdf) $F_N(x)$ with the theoretical
one $F(x)$ by computing their maximum deviations. When
$F(x)$ is continuous, we may form the following test statistics:
\begin{eqnarray}
K^{+} & = & \sqrt{N} \sup \{F_{N}(x)-F(x)\} ; \\
K^{-} & = & \sqrt{N} \sup \{F(x)-F_{N}(x)\} .
\end{eqnarray}
\noindent
$K^+$ measures the maximum deviation of $F_N(x)$ from
$F(x)$ when $F_N(x) > F(x)$ and $K^-$ measures the respective
quantity for $F_N(x) < F(x)$. These test statistics should be
distributed according to the KS distribution (with chosen
$N$), which is found in standard statistical textbooks.
Let $S$ then denote the KS statistic. As in the chi-square
test, a random number generator fails the KS test if under
the null hypothesis $H_0$ (that $K^+$ and $K^-$ obey the
KS distribution) any of the observed descriptive levels
$\delta^{+} = \mbox{Prob}( S \leq K^{+} \, | \, H_0)$ or
$\delta^{-} = \mbox{Prob}( S \leq K^{-} \, | \, H_0)$
is ``too small'' or ``too large''. Again, usually chosen
limits are 0.05 and 0.95, respectively. This kind of
test in which the empirical distribution is compared {\em once}
with the theoretical one is sometimes called a one-level
test \cite{Lec92}.

Unlike in the chi-square test, calculations of the KS test
statistics do not necessarily involve any approximation.
Therefore, the KS test can be reliably used with any value
of $N$. However, if $F_N(x)$
does not follow $F(x)$ precisely, fairly large values of $N$ would
confirm this difference better than small ones. On the other hand,
when a large number of observations are taken, locally
nonrandom behavior will tend to average out \cite{Knu81}. This
suggests using small values of $N$. This apparent contradiction is
avoided by making a compromise: choose $N$ fairly large and
apply another KS test to the previous test statistics $K^+$ and
$K^-$ by means of repeating the test $M \gg 1$ times. When
the empirical cdf of the test statistics $K^+$ and $K^-$ is
compared with the expected (approximate) distribution \cite{Knu81}
$F_M(x) = 1-\exp (-2x^2)(1-\frac{2x}{3\sqrt{M}}+{\cal O} (1/N))$, $x\geq 0$,
four new test variables $K^{++}$, $K^{+-}$, $K^{-+}$,
and $K^{--}$ are found. Then, the generator is considered to fail
the test if any of the respective descriptive levels
$\delta^{++}$, $\delta^{+-}$, $\delta^{-+}$, and $\delta^{--}$
are again ``too small'' or ``too large''. This is called
a {\em two-level} test \cite{Lec92}, which tends to detect
local correlations better than a one-level test \cite{Knu81}.

\bigskip
\section{$d$-tuple and rank tests}

In this Section, we will present brief descriptions of
the $d$-tuple and rank tests, which in this work are
the first tests in which the aforementioned chi-square and
KS tests are utilized. Both of these tests have been
designed to study bit level properties of random numbers.
For further details see Refs.
\cite{Alt88,Mar85,Mar85b,Vat93}.

\bigskip\bigskip
\noindent
{\bf $d$-tuple test}
\medskip

\noindent
The $d$-tuple test realized here is a modified version \cite{Alt88}
of the original \cite{Mar85}. We have extended the test \cite{Alt88}
further by improving its statistical accuracy by submitting the
empirical distribution of the test statistics \cite{Alt88} to a
Kolmogorov-Smirnov test. In the $d$-tuple test, we represent a random
integer $I_i$, $i=1,2,\ldots,n$, as a binary sequence of $b$ bits
$\tilde{b}_{i,j}, (j = 1,\ldots,b)$:
\begin{eqnarray}
I_{1} & = & \tilde{b}_{1,1} \tilde{b}_{1,2} \tilde{b}_{1,3} \cdots
	\tilde{b}_{1,b}, \nonumber \\
I_{2} & = & \tilde{b}_{2,1} \tilde{b}_{2,2} \tilde{b}_{2,3} \cdots
	\tilde{b}_{2,b}, \nonumber \\
      &   & \vdots \nonumber \\
I_{n} & = & \tilde{b}_{n,1} \tilde{b}_{n,2} \tilde{b}_{n,3} \cdots
	\tilde{b}_{n,b},
\label{Eq:d2test}
\end{eqnarray}
with an obvious choice for the parameter being $b=31$ (in testing RAN3,
we used $b=30$). Each of the binary sequences $I_i$ is divided into
subsequences of length $l$ which can be used to form $n$ new binary
sequences $I'_{i} = \tilde{b}_{i,1} \tilde{b}_{i,2} \ldots
\tilde{b}_{i,l}$.
These sequences are then joined into one more binary sequence of length
$d\times l$ in such a way that these final sequences
$\bar I_i$ partially overlap:
\begin{eqnarray}
\bar{I}_{1} & = & \tilde{b}_{1,1} \cdots \tilde{b}_{1,l}
	\tilde{b}_{2,1} \cdots \tilde{b}_{2,l}
	\cdots \tilde{b}_{d,1} \cdots \tilde{b}_{d,l}, \nonumber \\
\bar{I}_{2} & = & \tilde{b}_{2,1} \cdots \tilde{b}_{2,l}
	\tilde{b}_{3,1} \cdots \tilde{b}_{3,l}
	\cdots \tilde{b}_{d+1,1} \cdots \tilde{b}_{d+1,l}, \nonumber \\
            &   &   \vdots \nonumber \\
\bar{I}_{n} & = & \tilde{b}_{n,1} \cdots \tilde{b}_{n,l}
	\tilde{b}_{n+1,1} \cdots \tilde{b}_{n+1,l}
	\cdots \tilde{b}_{n+d-1,1} \cdots \tilde{b}_{n+d-1,l} .
\end{eqnarray}
Each of these new integers falls within $\bar I_i \in [0,2^{dl}-1]$.
In the test, the values $\bar I_i$ of the new random numbers are
calculated as well as the number of respective occurrences.
A statistic which follows the $\chi^2$ distribution can be
calculated although the subsequent sequences are correlated
\cite{Alt88}. The chi-square test is repeated $N$ times, and
the results are finally subjected to the KS test.

Based on studies of Altman \cite{Alt88} and Vattulainen
{\em et al.} \cite{Vat93}, the $d$-tuple test detects bit
level correlations effectively. In this work, however, our
main purpose is not to use this test in testing random
number generators but to compare its efficiency with
the cluster test.

\bigskip\bigskip
\noindent
{\bf Rank test}
\medskip

\noindent
The rank test is one of several tests proposed by Marsaglia
\cite{Mar85}, who developed it for studies of bit level
properties of random numbers. However, since our purpose in
this work is not to study generators by means of the rank
test but to compare its efficiency (to find known correlations)
with the cluster test, we have used a following version of
the rank test. Using the notation of Eq. (\ref{Eq:d2test}),
we first studied the bits of the first $w$ columns in the
random number sequence $I_i$, $i=1,2,\cdots,n \times v$ by
forming $n$ binary matrices of size $(v \times w)$. Since
consequtive samples were independent of each other (no
overlapping between consequtive matrices), we then calculated
$n$ values for the rank $R$, whose probability to equal one
of values $0,1,2,\ldots , \min(v,w)$ is \cite{Mar85,Mar85b}
\begin{equation}
p_{R} = 2^{R(v+w-R) - vw} \prod_{i=0}^{R-1}
\left[ \frac{(1-2^{i-w})(1-2^{i-v})}{ 1 - 2^{i-R}} \right].
\end{equation}
The $n$ values for the rank $R$ were then subjected to the
chi-square test. In a similar way we studied other
sequences of columns starting from bits $2,3,\ldots,b$, $b$
being the number of bits in a random word. Finally, this
procedure was repeated $N$ times in order to perform $b$ KS
tests. Bit number $i$ then failed, if the values for the
resulting descriptive levels of all the columns
$i-w+1,i-w+2, \cdots, i$ were smaller than 0.05 or
larger than 0.95.

Based on some bit level studies \cite{Vat93}, although less
efficient than the $d$-tuple test, also the rank
test seems to find correlations effectively.

\bigskip
\section{Presentation of new tests}

In the following, detailed descriptions of the tests developed
in this work will be given. The first two, the cluster test
and the autocorrelation test, are closely related to the
two-dimensional Ising model \cite{Bax82}. These tests are
respectively based on studies of the cluster size distribution
in a random lattice and on calculations of the integrated
autocorrelation times for certain physical quantities. Although
we apply these tests to the Ising model, they can be
generalized to other models and applications as well.
For example, our version of the cluster test is developed for
studies of random bits, but nothing restricts its use with
random words as well. Moreover, the idea of using
autocorrelation functions in testing of random numbers is
by no means restricted to the Ising model.

The next two test methods are related to random walks.
In the random walk test, we study random walks on a plane as
a function of the walk length. The $n$-block test is based on
the idea of renormalizing a sequence of uniformly distributed
random numbers. Despite its simplicity, the latter test
is especially effective in finding local correlations.

Finally, we present the condition number test which is based on
studies of Gaussian distributed random matrices. Moreover,
from a more general point of view tests on random matrices
are a particularly charming method, since random matrices
have applications in a wide area of science: nuclear physics
\cite{Mehta67}, chaotic systems \cite{Crisanti93}, computer
image generation \cite{Dia84}, and algorithm development
\cite{Ede_comm}, for example. For a recent review
on random matrices, see Ref. \cite{Crisanti93}.

\bigskip
\subsection{The cluster test}

There is a natural analogy between binary numbers and the Ising
model, which can be made use of in constructing a
{\it cluster test} in the following way. We take $i^{\rm th}$
bit from every successive number and put them on a
two-dimensional square lattice of size $L^2$. By identifying ones
and zeros with the ``up'' and ``down'' spins ${\cal S} = \pm 1$ of
the Ising model, the resulting configuration --- if truly random
--- should be one of the $2^{L^2}$ equally weighted configurations
corresponding to infinite temperature. The easiest quantity that
one can then compute from this analogy is the magnetization.
However, a better measure of {\it spatial} correlations can
be obtained if we study the distribution of connected spins, or
clusters of size $s$ on the lattice. The cluster size
distribution $\langle {\cal C}_s \rangle$ is given by \cite{Syk76}
\begin{equation}
\langle {\cal C}_s \rangle = s p^{s} D_{s}(p),
\end{equation}
in which $D_s (p)$'s are polynomials in $p=1/2$. The normalization
condition is $\sum_{s=1}^{\infty} \langle {\cal C}_s \rangle = 1$.
Enumeration of the polynomials $D_s (p)$ has been done up to
$s=17$ \cite{Syk76}, and they are listed in Appendix A.

The test procedure we have used is as follows. We first form
a $L^2$ lattice as above and enumerate all the clusters in it
\cite{Wan90} by using periodic boundary conditions in both
directions (\cite{Bin88} pp. 26-28).
For such a configuration we calculate the (unnormalized)
average size of clusters within $s = 1, 2, \ldots, 17$,
denoted as $S_{17}^{(k)}$. This procedure is repeated $N$ times
corresponding to configurational averaging, yielding
$S_{17} = \sum_{k=1}^{N} S_{17}^{(k)}/N$. The theoretical
counterpart of this quantity is given by
$ s_{17} = \sum_{s=1}^{17} s \langle {\cal C}_s \rangle$.
We also compute the empirical standard deviation
$\sigma_{17}$ of the quantities $S_{17}^{(k)}$.
For each $i^{\rm th}$ bit the test statistic chosen is:
\begin{equation}
g_i = \frac{S_{17} - s_{17}}{\sigma_{17}}.
\end{equation}
Using this statistic, the tests were performed comparatively
between several pseudorandom number generators, with results
from GGL assumed to be independent variables. Comparison of other
generators with GGL is justified since GGL has been shown to
have excellent properties on bit level \cite{Alt88,Vat93},
and as Fig. \ref{Fig:GGL_gauss} shows, the distribution for
$S_{17}^{(k)} - s_{17}$ is given by a Gaussian for GGL.
\begin{figure}[htb]
	\vspace{5.0cm}
\caption[The (unnormalized) distribution of GGL for
	 $D = S_{17}^{(k)} - s_{17}$ with 31\,000
	 independent samples]
	{The (unnormalized) distribution (circles) of GGL for
	 $D = S_{17}^{(k)} - s_{17}$ with 31\,000
	 independent samples. Although this Figure is made for
	 illustrated purposes only, we may still observe that
	 the distribution clearly follows Gaussian behavior
	 (indicated with a solid line). \label{Fig:GGL_gauss}}
\end{figure}
Therefore, the mean value of $g_i$ over all the 31 bits of GGL, denoted
as $g_{\rm GGL}$ and the corresponding standard deviation
$\sigma_{\rm GGL}$ were computed and the results for all other
generators were compared with these values using
\begin{equation}
\label{Eq:gidot}
g_i'=  \frac{\vert g_i - g_{\rm GGL} \vert }{\sigma_{\rm GGL}}.
\end{equation}
The bit $i$ in question failed the test if $g_i'$ was
consequtively greater than three in two separate tests.

We also considered other similar choices for the test
parameters such as using the maximum value of $g_i$ over
all the 31 bits of GGL instead of $g_{GGL}$, and then
performing the analysis as above. The results of
this approach were consistent with Eq. (\ref{Eq:gidot})
(results of bits 7 and 12 of RAND being the only exceptions).

\bigskip
\subsection{Autocorrelation test}

In {\em the autocorrelation test} we consider the autocorrelation
function $C_A$ for some physical quantity $A$. Then, we calculate
the integrated autocorrelation time $\tau_A$ of $C_A$.
Our approach follows the procedure given in Ref. \cite{Wol89b},
whose main details are given below.

The autocorrelation function is defined as
\begin{equation}
C_A(t) = \frac{\langle A(t_0)A(t_0+t) \rangle
- \langle A(t_0)\rangle^2 }{\langle A(t_0)^2\rangle
 - \langle A(t_0)\rangle^2},
\end{equation}
where $t$ denotes time.
In order to calculate an estimator $\tau_A (W)$ for
the integrated autocorrelation time $\tau_A$, a truncation
window $W$ is used:
\begin{equation}
\label{Eq:tausum}
\tau_A (W) = \frac{1}{2} + \sum_{t=1}^{W-1} C_A(t) + R(W),
\end{equation}
with the remainder
\begin{equation}
R(W) = \frac{C_A(W)}{1 - \gamma (W)}
\end{equation}
and
\begin{equation}
\gamma (W) = \frac{C_A(W)}{C_A(W-1)}.
\end{equation}
The convergence of $\tau_A (W)$ must be checked as a function of
the window size $W$. Since noisy contributions from large separations
appear after some value $W_{n}$, the estimate $\tau_A$ is found
by averaging $\tau_A(W)$ between $W_c$ and $W_n$, in which
$W_c < W_n$ denotes the value for which Eq. (\ref{Eq:tausum})
first converges. An illustration of this procedure is given
in Fig. \ref{Fig:GGL_susc_wolff}.
\begin{figure}[htb]
	\vspace{5.0cm}
\caption[The integrated autocorrelation time $\tau_{E}$ of
	energy $E$ in the autocorrelation test when RAND has
	been employed.]
	{The integrated autocorrelation time $\tau_{E}$ of
	energy $E$, when RAND has been employed.
	The error starts to increase after $W_n = 18$. The
	error bars in the cases of $W=19$ and $W=21$ extend
	beyond the graph. \label{Fig:GGL_susc_wolff}}
\end{figure}
The error estimate for $\tau_A(W)$ is given in Ref. \cite{Wol89b}.

In this work, we considered the two-dimensional Ising model.
The simulations were carried out on a square lattice
with the Wolff algorithm \cite{Wol89a} at the well known
critical coupling $K_c = \frac{1}{2} \ln (1+\sqrt{2})$.
The linear size of the system was $L=16$.
Our implementation of the single cluster search algorithm
followed Ref. \cite{Wan90}, and the measurements for the
calculated quantities the energy $E$, the magnetic susceptibility
$\hat\chi$ \cite{Wol89b}, and the (normalized) size of the
flipped clusters $c$ were separated by one single cluster
update only. Then, by following the procedure given above
we calculated the corresponding {\it integrated
autocorrelation times} $\tilde{\tau}_E$, $\tilde{\tau}_{\hat \chi}$,
and $\tilde{\tau}_c$ by first calculating their autocorrelation
functions $C_E, C_{\hat \chi}$, and $C_c$. Finally, these
estimates for the integrated autocorrelation times were
scaled to the time unit of one Monte Carlo step;
{\em i.e.} every spin on the lattice is updated once on the
average. Therefore, the final results are
$\tau_A = \tilde{\tau}_A \langle c \rangle$ \cite{Wol89b},
in which $A$ is one of the quantities $E$, $\hat\chi$, and $c$.

In the case of the two-dimensional Ising model, the exact value
for the energy $E = 1.45312$ \cite{Ferdi69} is known, which
allows a comparison between results from
different pseudorandom number generators. For other
quantities, the test provides us with information on the
relative performance of the random number generators.
Here we assumed the results from GGL and RANMAR to be
correct. This assumption is justified because of their
results for the energy $E$, which in our simulations
were correct within error limits.

\bigskip
\subsection{Random walk test}

Random walks have applications in a very wide area of
science: linear polymer molecules may be modelled with
self-avoiding random walks \cite{Sok94}, (non-biased)
random walks are used in studies of {\em e.g.} diffusion
limited aggregation \cite{Vic89}, and so on. Therefore,
the idea of using random walks as a test for randomness
seems a very interesting and useful idea.

In {\it the random walk test}, we consider random walks on a
two-dimensional lattice, which is divided into four equal
blocks, each of which has an equal probability to contain
the random walker after a walk of length $n$. The test is
performed $N$ times, and the number of occurrences in each
of the four blocks is compared to the expected value of $N/4$,
using the chi-square test with three degrees of freedom. The
generator fails if the $\chi^2$ value exceeds 7.815 in at
least two out of three independent runs. This should occur
with a probability of only about 3/400.

For the purpose of completeness, let us mention that
other random walk tests have been proposed by Binder
and Heermann (\cite{Bin88} p. 76-80) and Ziff \cite{Zif92_pre}.
The former is based on the idea of studying
the average end-to-end distance which should be a linear
function of the walk length $n$. The test proposed by Ziff
is based on random walks in a two-dimensional square lattice,
where the random walker starts from one corner and
heads towards the opposite one. At every step it may turn
either left or right, unless it enters a previously visited
site in which case it is forced to turn so as not to retrace
its path. Therefore, eventually it hits one of the two opposite
boundaries, which should occur with an equal probability.

\bigskip
\subsection{$n$-block test}

{\em The $n$-block test} is a simplified version of our
random walk test, being basically a random walk test
in one dimension. In this test we take a sequence
$\{x_1,x_2,\ldots,x_n\}$ of uniformly distributed random numbers
$0\leq x_i < 1$, whose average $\bar{x}$ is calculated. If
$\bar{x} \geq 1/2$, we choose $y_i = 1$; otherwise $y_i = 0$.
This is repeated $N$ times. We then perform the chi-square test on
variables $y_i$ with one degree of freedom. Each test is repeated
three times, and the generator fails the test with fixed $n$ if
at least two out of three $\chi^2$ values exceed 3.841, which
should occur with a probability of about $3/400$.

In Ref. \cite{Gra93b}, Grassberger has proposed a somewhat
unspecified ``block'' test to study the range of correlations
for LF$(17,5,+)$.

\bigskip
\subsection{Condition number test}

An interesting method for studies of pseudorandom number generators
is by means of random matrices. We have constructed one test
of this kind, concentrating on distributions of condition
numbers in Gaussian distributed random matrices. This
{\em condition number test} is based on the theoretical work
of Edelman \cite{Ede88}.

Consider a real $m \times 2$ matrix $B$ with elements from a
standard normal (Gaussian) distribution. We define a Wishart
matrix $W = BB^{T}$, and calculate its eigenvalues $\lambda_{max}$
and $\lambda_{min}$ ($\lambda_{max} \geq \lambda_{min} \geq 0$)
and the 2-norm condition number
$\kappa = \sqrt{\lambda_{max} / \lambda_{min}}$ of $B$.
The probability distribution function of $\kappa$
for such a Gaussian matrix is \cite{Ede88}
\begin{equation}
f(\kappa) = (m-1) 2^{m-1} \frac{\kappa^{2} - 1}
			  {(\kappa^{2} + 1)^m} \ \kappa^{m-2},
\end{equation}
and its cumulative distribution function (cdf) is given by
\begin{eqnarray}
\label{Eq:Cumu}
F(b) 	& = & \int_{1}^{b} f(\kappa) \ d\kappa \\ \nonumber
	& = & 1 - \left( \frac{2b}{1 + b^{2}} \right)^{m-1}, \,\,\,
	      b \geq 1.
\end{eqnarray}
In the condition number test we proceed as follows. First,
by using a pseudorandom number generator we form a Gaussian
distributed $m \times 2$ matrix $B$ and calculate its condition
number $\kappa$. This is repeated $N$ times. Then, we compare
the empirical cdf with the one given by Eq. (\ref{Eq:Cumu})
by using a Kolmogorov-Smirnov test. At this point, two test
statistics $K^+$ and $K^-$ are found. Then, the previous
procedure is repeated $M$ times, and another KS test is
performed for the cdf's of test statistics $K^+$ and $K^-$.
As a result, we find the final test variables $K^{++}$,
$K^{+-}$, $K^{-+}$, and $K^{--}$. In this work, we considered
the generator to fail the test
if any of the respective descriptive levels $\delta^{++}$,
$\delta^{+-}$, $\delta^{-+}$, and $\delta^{--}$ were less than
0.05 or larger than 0.95. In other words, a failure occurred if
the empirical cdf followed too closely or was too far from the
theoretical one.

Other tests based on the properties of random matrices can also
be constructed. For example, one possibility is to use the expected
number of real eigenvalues in a Gaussian distributed
random matrix \cite{Ede92p}. However, our experience
shows that this test is not very
efficient\footnote{\protect\samepage{We performed
few bit level tests based on this idea. In the tests, we formed
31-bit integers of bits $i$ ($i=1,2,\ldots,31$) in the output
of a pseudorandom number generator, and transformed them into
Gaussian distributed random variables by means of the Box-Muller
method \protect\cite{Box58}. With matrices of size
$50 \times 50$ no correlations with any generator studied were found.}}.
Other ideas consist of using distributions of smallest eigenvalues
\cite{Ede91} or of scaled condition numbers \cite{Ede92} as a
basis for empirical tests. Finally, Marsaglia \cite{Mar85,Mar85b}
has proposed a test based on the calculation of the rank of
a random matrix (cf. Section 4.4).

In testing Gaussian distributed random variables one important
factor must be taken into account: since pseudorandom number
generators usually produce uniformly distributed random variables,
some method must be used to transform them into Gaussian distributed
random variables. This raises an obvious question: are we
studying the goodness of the transformation method or the
goodness of uniformly distributed random variables? In order
to clarify this question, we have studied few different
transformation methods, and chosen the best one for further
condition number tests on pseudorandom number generators.
In the following Section, these transformation methods
will be introduced.

\bigskip
\section{Generation methods for Gaussian distributed random variables}

In addition to uniformly distributed random numbers, variables
from numerous other distributions are often needed. One of the
most frequently used distributions is the normal (Gaussian)
distribution, because of its importance in the field of
statistics and probability theory \cite{Tez91a}, for example.
In this Section, we consider some generation methods for
Gaussian distributed random variables. The algorithms of
these methods are given below, where we assume that all
$x_i$, $i=1,2,\ldots$ are identically and independently
distributed random variables from the $U(0,1)$ distribution
(the uniform distribution between zero and one);
for other methods see Refs. \cite{Bra83} pp. 134-179,
\cite{Knu81} pp. 114-133, and \cite{Rub81} pp. 38-107.

One of the most common techniques in generation of such
variables $X_i$, $i=1,2$, is the Box-Muller algorithm \cite{Box58}:
\begin{table}[htb]
\small\tt
\begin{tabular}{l l l l l}
 & & & $\triangleright \, 1$ & Generate $x_1$ and $x_2$. \\
 & & & $\triangleright \, 2$ & Let $X_1 = \sqrt{-2 \ln x_1}
	\cos (2\pi x_2)$
		and $X_2 = \sqrt{-2 \ln x_1} \sin (2\pi x_2)$. \\
\end{tabular}
\end{table}

\noindent
Despite its popularity, this method has several drawbacks. First,
since the Box-Muller method uses several multiplications,
one trigonometric and logarithmic function, and a square
root per a single random variable, it is very slow.
In addition, the tail distribution differs markedly from
the true distribution, when LCG or Tausworthe generators
are used \cite{Tez91a}. Therefore, in this work we have considered
three other methods instead of the Box-Muller
algorithm. The chosen methods include the so called
Marsaglia's polar method \cite{Rip83}, the ratio method
\cite{Rip83}, and a method based on the central limit
theorem \cite{Rip88}.

The algorithm of {\em Marsaglia's polar method} \cite{Rip83} is
basically a Box-Muller algorithm, in which the sine and cosine
computations have been eliminated by a rejection technique:
\begin{table}[htb]
\small\tt
\begin{tabular}{l l l l l}
 & & & $\triangleright \, 1$ & 	Generate $x_1$, $x_2$, and
				$y_i = 2 x_i - 1$, $i=1,2$. \\
 & & & $\triangleright \, 2$ & 	Let $z = y_1^2 + y_2^2$. If $z>1$
				goto 1. \\
 & & & $\triangleright \, 3$ & 	Let $C = \sqrt{(-2 \ln z)/z}$,
				$X_1 = C x_1$, and $X_2 = C x_2$. \\
\end{tabular}
\end{table}

\noindent
This method generates two Gaussian distributed random variables
$X_1$ and $X_2$ at a time, and utilizes only one logarithm and
a square root. The price it must pay is the efficiency, since
only $\pi / 4 \approx 0.786$ normal deviates are produced per
one uniform random number.

One way to write {\em the ratio algorithm} for the normal
distribution \cite{Rip83} is:
\begin{table}[htb]
\small\tt
\begin{tabular}{l l l l l}
 & & & $\triangleright \, 1$ & 	Generate $x_1$, $x_2$, and
				$y = \sqrt{2/e}\,(2 x_2 - 1)$. \\
 & & & $\triangleright \, 2$ & 	Form $X = y/x_1$, $z = X^2 / 4$. \\
 & & & $\triangleright \, 3$ & 	If $z > - \ln x_1$ \,\, go to 1,
				otherwise return $X$. \\
\end{tabular}
\end{table}

\noindent
Although this method does not calculate the square root,
it is slower than Marsaglia's method since it generates
only one Gaussian distributed random variable $X$ with an
approximate efficiency of 0.366.

The third method considered in this work relies on the central
limit theorem \cite{Rip88}, in which $n$ uniform random deviates
$x_i$ are used to form one Gaussian distributed random
variable $X$:
\begin{table}[htb]
\small\tt
\begin{tabular}{l l l l l}
 & & & $\triangleright \, 1$ & Generate $x_1, x_2, \ldots, x_{n}$. \\
 & & & $\triangleright \, 2$ & Form
		$X = (\sum_{i=1}^{n} x_i - n/2) \sqrt{12/n}$. \\
\end{tabular}
\end{table}

\noindent
The main advantage of this method is its speed, if small values
of $n$ are used. Nevertheless, it is a well known fact that this
method converges fairly slowly towards the Gaussian distribution,
and therefore small values of $n$ should be avoided. In the
literature, however, this method is still often mentioned,
usually with the choice $n = 12$ \cite{Bra83,Rub81}.

In testing the generation methods for Gaussian distributed
random variables we used a simplified version of the condition
number test. First, by using RANMAR we formed a Gaussian
distributed $m \times 2$ matrix $B$ and calculated its condition
number $\kappa$. This was repeated $N$ times. Then, we compared
the empirical cdf with the expected one (Eq. (\ref{Eq:Cumu}))
by using a Kolmogorov-Smirnov test, and found two test
statistics $K^+$ and $K^-$ and their respective descriptive
levels $\delta^{+}$ and $\delta^{-}$ ($\delta^{+}, \delta^{-}
\in [0,1)$).
When this procedure was repeated $M$ times, we calculated the total
number $G$ of descriptive levels which were less than 0.01 or larger
than 0.99. Since the maximum value of $G$ is $2M$, we finally
obtained the test statistic $\epsilon = G/2M$. The tested
generation method was considered good, if $\epsilon$ was
$0.02 \pm 0.01$.


\chapter{Results}

In this Chapter, results of the tests described in Section 4.5
will be given. Most Tables and Figures are given in the
context of the text, although for purposes of clarity two
Tables containing the exact values of the test statistics are
given in the Appendices. Moreover, in addition to reporting
the results and discussing some open problems our results
clarify, the properties of the tests developed in
this work will be discussed in some extent.

Before proceeding to the results, however, some technical
facts must be given. In this test program, the initial seed
values for pseudorandom number generators are chosen from
the set $\{ 12345, 667790, 14159, 1415926535, 97766\}$.
The tests have been performed on computers of DEC 3000 AXP
series and Convex C3840. In some tests routines of the
NAG library have been used, other code being written during
this work. In the cases where test codes have been used on
different machines, portability has been checked.
Parallelization has not been utilized.

\pagebreak

\section{Cluster test}

In this Section, we will study bit level correlations of
some commonly used pseudorandom number generators by means
of three tests: the $d$-tuple test \cite{Alt88,Mar85},
the rank test \cite{Mar85,Mar85b},
and {\em the cluster test}. First, we study their efficiency
in finding periodic correlations in random bit sequences, and by
this means compare them. Then, we apply the cluster test to
several generators and compare these results with previous
results of the $d$-tuple and rank tests.

\bigskip
\subsection{Studies on efficiency}

The $d$-tuple and rank tests have been shown to find bit level
correlations
efficiently \cite{Vat93}. In order to determine the quantitative
effectiveness of the tests, we first studied their ability to observe
correlations inserted into the output of GGL, which has passed
the bit level tests in Ref. \cite{Vat93}. The correlations were
inserted periodically by setting the $i^{\rm th}$ bit
($i=1,2,\ldots,31$) of every $\xi^{\rm th}$ number always equal
to one. By systematically varying $\xi$, we could then find the
maximum approximate distance $\xi_c$ within which the $d$-tuple
and rank tests can detect periodic correlations. The $d$-tuple
test was repeated three times with parameters $d=l=3$, $n=5000$
and $N=1000$. Accordingly, the rank test was repeated three times
with parameters $v=w=2$, $n=1000$, and $N=1000$. The
results are shown in Table \ref{Tab:d2p}, where the parameter
$p$ gives the probability of observing correlations.
\begin{table}[htb]
\small \centering
\begin{tabular}{| c | c | c | c | c | c | c | c | c | c | c |}
\hline\hline
\multicolumn{11}{| c |}{Results of the $d$-tuple test} \\
\hline
$\xi$   & 40    & 43    & 52    & 60    & 70
        & 80    & 90    & 100   & 110   & 120   \\ \hline
$p$     & 1.000 & 0.889 & 0.778 & 0.333 & 0.667
        & 0.222 & 0.333 & 0.111 & 0.222 & 0.000 \\
\hline\hline
\multicolumn{11}{| c |}{Results of the rank test} \\
\hline
$\xi$   & 40    & 45    & 50    & 60    & 70
        & 80    & 90    & 100   & 110   & 120   \\ \hline
$p$     & 1.000 & .667  & 0.667 & 0.333 & 0.167
        & 0.50  & 0.167 & 0.000 & 0.333 & 0.000 \\ \hline\hline
\end{tabular}
\caption[Results of the $d$-tuple and rank tests
	 with inserted correlations in the bits.]
	{Results of the $d$-tuple (above) and rank tests (below)
	 with inserted correlations in the bits, with a period of
	 $\xi$. The probability for the tests to observe correlations
	 is denoted by $p$, which for the $d$-tuple and rank tests
	 equals one up to $\xi_c \approx 43$. \label{Tab:d2p}}
\end{table}
Thus, the $d$-tuple test can detect periodic correlations up to
$\xi_c \approx 43$ bits apart. The same test was repeated with $d=9$
and $l=1$ to consider single bits only, which gave $\xi_c \approx 50$.
Similar systematic tests for the rank test also gave a result
$\xi_c \approx 43$, but in the range $40-70$ for $\xi$ the $d$-tuple
test was found to be slightly more efficient. Hence, based on
our results the rank test is slightly inferior to
the $d$-tuple test.

The effectiveness of the cluster test was first scrutinized
by inserting periodic correlations as in the previous cases.
We chose $L=200$, $N = 10\,000$ and the study was
repeated for all values of $\xi = 1, 2, \ldots, L$. The results
are shown in Table \ref{Tab:Frame}, where filled squares denote
distances where correlations were detected.
\begin{table}[htb]
	\vspace{10cm}
\caption[Results of the cluster test with correlations in the bits.]
	{Results of the cluster test with correlations in the
	 bits, with a period of $\xi$ from one to 200. Black
	 squares denote corresponding distances at which
	 correlations were found as explained in the text.
	 \label{Tab:Frame}}
\end{table}
With this choice of parameters, the
cluster test is able to find all periodic correlations up to
$\xi_c \approx 60$. Moreover, due to the periodic boundary
conditions (cf. Section 4.5.1) further correlations with larger
values of $\xi$ were also observed. This shows that the cluster
test performs somewhat better than either the $d$-tuple
or rank tests.

\bigskip
\subsection{Testing generators using the cluster test}

Next, we subjected each bit of the random number generators
to the cluster test. It was repeated twice with parameters
$L=200$ and $N=10\,000$. To confirm exclusion of finite-size
effects additional tests with $L=500$ were carried out. They
gave consistent results with $L=200$. Results are summarized
in Table \ref{Tab:Cluster}, where results of the previous
$d$-tuple and rank tests from Ref. \cite{Vat93} have also
been included.
\begin{table}[htb]
\small \centering
\begin{tabular}{| l | c | c | c | c |}
\hline\hline
\multicolumn{1}{| c |}{Random} & \multicolumn{4}{ c |}{Failing bits} \\
\cline{2-5}
\multicolumn{1}{| c |} {number}
& \multicolumn{1}{c |}{\makebox[2.9cm]{Cluster test}}
& \multicolumn{1}{c |}{\makebox[2.9cm]{$d$-tuple test}}
& \multicolumn{1}{c |}{\makebox[2.9cm]{Rank test}}
& \multicolumn{1}{c |}{\makebox[2.9cm]{Equidistribution}} \\
\multicolumn{1}{| c |}{generator}
& \multicolumn{1}{c |}{}        & \multicolumn{1}{c |}{}
& \multicolumn{1}{c |}{}        & \multicolumn{1}{c |}{of bits}
\\ \hline\hline
GGL     & none                  & none                  & none
        & none  \\
R250    & none                  & none                  & none
        & none  \\
R1279   & none                  & none                  & none
        & none  \\
RANMAR  & $25-31$               & $25-31$               & $25-31$
        & $25-31$ \\
RAN3    & $1-4$, $25-30$        & $1-5$, $25-30$        & $1-5$, $26-30$
        & $1-11$, $24-30$ \\
RAND    & $7-31$                & $13-31$               & $18-31$
        & $22-31$ \\
\hline\hline
\end{tabular}
\caption[Results of the cluster test.]
	{Results of the cluster test ($k=1$), where bit number one
	 denotes the most significant bit. $d$-tuple and rank test
	 results are from Ref. \protect\cite{Vat93}. The last
	 column denotes bits which fail in testing the
	 equidistribution of ones. \label{Tab:Cluster}}
\end{table}
More detailed results of the cluster test are given
in Appendices B and C. Although more powerful than the other
methods, the cluster test still reveals no discernible
correlations for either GGL, R250 or R1279. For RANMAR and
RAN3, the cluster test gives results consistent with Ref.
\cite{Vat93}, but for RAND additional correlations are revealed
in bits $7 - 12$, in contrast of passing the $d$-tuple test
\cite{Vat93}. According to the results of RAND, the cluster
test is very effective in locating periodic correlations, since
the period of bit number 8 of RAND is as large as $2^{24}$
(see discussion of RAND on page \pageref{rand}).

For completeness, we also tested {\em the equidistribution of bits}.
The bits failed the test if the deviation from the expected
number of ones ({\em i.e.} $L^2$/2) consequtively exceeded
three times the standard deviation of the binomial distribution
with $M$ samples. The test was repeated twice with
$M = 4 \times 10^8$ and its results are also shown in Table
\ref{Tab:Cluster}. No correlations were found for GGL, R250,
or R1279. Surprisingly, however, this rather simple test revealed
that the first 11 bits of RAN3 fail (with standard deviations
larger than 6.7) although only the first four or five bits fail
in the other tests. On the other hand, for RAND only bits
$22 - 31$ failed, which produced an exact $50 - 50$
distribution of zeros and ones.

In conclusion, the cluster test in the form presented here,
is very well suited for detection of correlations on bit level,
being especially effective for periodic correlations.

\bigskip
\section{Autocorrelation test}

{\em The autocorrelation test} was carried out with two sets
of parameters. First, $10\,000$ Monte Carlo steps (MCS's) were
performed to equilibrate the system starting from a random initial
state, and then $N=10^7$ samples were taken to test most of the
generators once. One MCS denotes updating of each lattice site
once on the average. In the second set, $100\,000$ MCS's  were
followed by $N=10^8$ samples to test some generators more
extensively. In the results considered here, the linear size
of the system was $L=16$.

A summary of the results in Table \ref{Tab:Auto} shows that based on
this test, the generators can be classified into two categories.
\begin{table}[thp]
	\vspace{5.0cm}
\caption[Results of simulations for the Ising model at criticality with
	the Wolff algorithm.]
	{Results of simulations for the Ising model at criticality with
	the Wolff algorithm. The number of samples is denoted by $N$
	and the values of lags $q_i$ are given where needed. The value
	of the decimation parameter $k$ is one unless stated otherwise.
	The errors shown correspond to $\sigma$ \protect\cite{Wol89b};
	1.447(5) denotes $1.447 \pm 0.005$ etc. .
	The most erroneous results are in boldface. See text for details.
	\label{Tab:Auto}}
\end{table}
First, let us consider results with $N=10^7$ samples.
For the energy $\langle E \rangle$, deviations from
the exact result of $\langle E \rangle = 1.45312$ \cite{Ferdi69}
for R31, R250, R521, and RAN3 are much larger than $3\sigma$ in which
$\sigma$ is the standard deviation \cite{Wol89b}. In particular,
the average size of flipped clusters $\langle c \rangle$ is very
sensitive to correlations in random number sequences, since
in the erroneous cases it is clearly biased. Most striking, however,
is the behavior of the integrated autocorrelation times $\tau$.
For generators, which show no significant deviations in
$\langle E \rangle$, $\langle \hat \chi \rangle$, or
$\langle c \rangle$, results for the $\tau$'s agree well with
each other. However, for R31 and R250, the integrated autocorrelation
times show errors of about 8\% compared with results of GGL and
RANMAR. We thus propose these quantities as
particularly sensitive measures of correlations in pseudorandom
number sequences.

Another important point is the behavior of R31 compared with
ZIFF31 and PENTA31. Though R31 clearly fails these autocorrelation
tests, its 5-decimated sequence ZIFF31 and a generator PENTA31
based on a primitive pentanomial $x^{31} + x^{23} + x^{11} + x^{9} + 1$
give correct results within error limits. This is further emphasized
in studies with $N=10^8$ samples, where R89 fails whereas ZIFF89
and PENTA89 give compatible results with RANMAR. Therefore, these
results clearly indicate that $k$-decimation of GFSR generators
with two lags and primitive pentanomials generate (in statistical
sense) more reliable sequences than GFSR generators based on
two lags only.

To compare our results with those of Refs. \cite{Fer92,Sel93} we also
used the autocorrelation time test to further study the decimation of
the output of R250, {\em i.e.} we took every $k^{\rm th}$ number of
the pseudorandom number sequence. For $k = \{3,5,6,7,9,10,11,12,24,48\}$,
the correlations vanish in agreement with Ref. \cite{Fer92} ($k=5$)
and Ref. \cite{Sel93} ($k=3,5$). On the other hand, for $k=2^m$
with $m=\{0,1,2,3,4,5,6,7,8\}$, the sequences fail. These findings
agree with the theoretical result of Golomb (\cite{Gol82} pp. 78-79)
who showed that the decimation of a maximum-length GFSR sequence
by powers of two results in equivalent sequences.

Our results of the autocorrelation test are in agreement with
observations made by various other authors \cite{Cod93,Fer92,Sel93},
who have also studied the two-dimensional Ising model.
Moreover, the errors in the average cluster sizes show that
the origin of errors observed in these references lies in
local correlations present in the cluster formation process
of the Wolff algorithm. The main advantage of our approach
is the use of integrated autocorrelation times as measures for
nonrandomness, since in the erroneous cases the errors are of
the order of several percents, whereas for other quantities
such as the energy the error is much smaller. Due to the fact
that this test is not restricted to the Ising model only, its
use in other problems might be very fruitful.

\bigskip
\section{Random walk test}

Although errors in the average cluster sizes for some
of the GFSR generators in the autocorrelation test suggest that
the correlations are within the ${\cal O}(L^2)$ successive
pseudorandom numbers, this result does not give a quantitative
support for the argument that the correlation length
equals the longer lag $p$ (cf. Section 3.1). Therefore, to quantify
the range of correlations empirically, we have performed {\it the
random walk test}.

First, we studied a group of generators with the walk length
$n=1000$. These results are presented in Table \ref{Tab:Rwalks},
\begin{table}[htb]
\small \centering
\begin{tabular}{| l | l | r | c | c |}
\hline\hline
\multicolumn{1}{| c |}{RNG} &
\multicolumn{1}{  c |}{$k$} &
\multicolumn{1}{  c |}{$q_i$} &
\multicolumn{1}{  c |}{$\chi^2$ values} &
\multicolumn{1}{  c |}{Result} \\
\hline
R31             &               & 3     	& 4094, 4105, 4300    &
	FAIL \\
R250            & 1,2,64        & 103   	& $396.4 - 539.8$     &
	FAIL \\
R521            & 1,2,64        & 168   	& $49.01 - 79.16$     &
	FAIL \\
RAN3            &               &       	& 40.01, 42.99, 44.53 &
	FAIL \\
\hline
R250            & 3             & 103   	& 0.301, 0.873, 1.024 &
	PASS \\
R521            & 3             & 168   	& 1.249, 1.352, 1.735 &
	PASS \\
R1279           & 1,2,3,64      & 418   	& $0.709 - 9.372$     &
	PASS \\
R4423           &               & 2098  	& 0.621, 1.226, 8.217 &
	PASS \\
PENTA31		& 		& 23,11,9	& 0.685, 1.587, 2.363 &
	PASS \\
ZIFF31		& 		& 13,8,3	& 2.352, 2.367, 2.632 &
	PASS \\
RAND		& 		& 		& 0.304, 0.640, 4.063 &
	PASS \\
RAN3            & 2,3           &       	& $0.033 - 6.877$     &
	PASS \\
GGL             &               &       	& 0.090, 0.459, 1.981 &
	PASS \\
RANMAR          &               &       	& 0.293, 1.944, 3.187 &
	PASS \\
\hline\hline
\end{tabular}
\caption[Results of the random walk test with parameters
	$n=1000$ and $N=10^6$.]
	{Results of the random walk test with $N=10^6$ samples.
	The parameter $k$ equals one unless stated otherwise.
	The fourth column indicates the $\chi^2$ values in three
	independent tests, or the range of the $\chi^2$ values
	when results with more than one value of $k$ are included
	on the same line. The classification of the generators
	is based on the failing criterion given in Section 4.5.3:
	a generator fails the test if the $\chi^2$ value exceeds
	7.815 in at least two out of three independent runs.
	See text for details. \label{Tab:Rwalks}}
\end{table}
and they are in agreement with the autocorrelation test.
No correlations for either
GGL, RAND or RANMAR were observed. R250 and R521 pass the test with
$k=3$, but fail with $k=\{1,2,2^6\}$, whereas R1279 passes
with all $k$'s tested. The failure of RAN3 with $k=1$ is consistent
with results of previous tests \cite{Vat93} and the autocorrelation
test. It is notable that all the failures in this test were very
clear, since  even the smallest $\chi^2$ values exceeded 40.
However, RAN3 passed the test when every second or third
number was used.

The main difference between the failing generators R250 and R521
(with $k=1$) and the successful ones R1279 and R4423 lies in the
lag parameter $p$ which is less than $n$ for the former and
larger than $n$ for the latter. We studied this for various values
of $p$ with the random walk test by locating the approximate
value $n_c$, above which the generators fail. The test was
performed for R31, R250, R521 and R1279 with $N=10^6$ samples.
The results for $n_c$ were $32 \pm 1$, $280 \pm 5$, $590 \pm 5$,
and $1515 \pm 5$, respectively, in which the error estimate is the
largest distance between samples close to $n_c$. For the purpose
of illustration, in Fig. \ref{Fig:R31_250} we show an example
of the $\chi^2$ values for R31 and R250 as a function of the
walk length $n$.
\begin{figure}[htbp]
\vspace{5.0cm}
\caption[The $\chi^2$ values for R31 and R250 in the random walk test
	as a function of walk length $n$.]
	{The $\chi^2$ values for R31 (inner figure) and R250
	($k=1$) in the random walk test as a function of walk
	length $n$, when $N=10^6$ samples have been taken.
	Three independent runs in both cases are denoted by
	different symbols. The horizontal lines denote
	$\chi^2=7.815$. \label{Fig:R31_250}}
\end{figure}

As another illustration of nonrandom behavior
we consider the probability distribution functions (pdf's)
for the final position of a random walker after $n$ steps
for GGL and R250. Since GGL passed the random walk test,
we assume that its pdf is approximately correct. Then,
we plot the difference between pdf's of R250 and GGL,
as has been done in Fig. \ref{Fig:XYZ}.
\begin{figure}[htbp]
\vspace{5.0cm}
\caption[The deviation $p_{\mbox{\scriptsize diff}}$ between (normalized)
	probability distribution functions of R250 and GGL as a function
	of lattice site.]
	{The deviation $p_{\mbox{\scriptsize diff}}$ between
	(normalized) probability distribution functions
	of R250 ($k=1$) and GGL as a function of lattice site, when
	the pdf of GGL was subtracted from the one of R250. The
	starting point of the random walk is (0,0). In
	this study we used $n=1000$ and $N=10^6$.
	\label{Fig:XYZ}}
\end{figure}
We note that the probability distribution of R250 deviates
significantly from that of GGL, since instead of
uniform noise we notice two clear peaks. As a matter of
fact, the deviation is very significant, since the relative
error of R250 compared to GGL is about 6.5 \% at these two
peaks observed in the figure.

We also studied generators, which are based on primitive pentanomials
or decimation of GFSR generators with two lags. As Table \ref{Tab:Rwalks}
indicates, PENTA31 and ZIFF31 pass the random walk test with $N=10^6$
samples. In these cases, studies to locate $n_c$ were inconclusive, since
a small period of these generators did not allow testing them with
more than $10^7$ samples. Therefore, similar studies for PENTA89 and
ZIFF89 with $N=10^8$ samples were carried out, these results being
summarized in Table \ref{Tab:RW2d}.
\begin{table}[htbp]
\small \centering
\begin{tabular}{| r | l | l | l | l | l | l |}
\hline\hline
\multicolumn{1}{|  c |}{$n$} &
\multicolumn{3}{   c |}{$\chi^2$ for ZIFF89} &
\multicolumn{3}{   c |}{$\chi^2$ for PENTA89} \\
\hline
85  & 7.785	  & 7.544	& {\bf 8.131} & 0.427	    & 1.132
	      & 1.711 \\
90  & 2.050       & 3.716       & 2.165       & 1.338       & 3.874
	      & 1.509 \\
95  & {\bf 10.93} & 6.642       & 3.895       & 1.335       & 3.563
	      & 3.422 \\
100 & {\bf 9.130} & 5.910       & {\bf 8.770} & 3.867       & 5.611
	      & 4.350 \\
200 & {\bf 8.632} & {\bf 15.76} & {\bf 13.61} & {\bf 25.02} &
	{\bf 18.48} & {\bf 17.56} \\
500 & 1.007	  & 6.173 	& {\bf 9.822} & {\bf 34.90} &
	{\bf 39.74} & {\bf 39.51} \\
\hline\hline
\end{tabular}
\caption[Some results of the random walk test with $N=10^8$ samples
	for PENTA89 and ZIFF89.]
	{Some results of the random walk test with $N=10^8$ samples
	for PENTA89 and ZIFF89. For both generators three independent
	runs have been performed. Failing results are shown with
	bold type. See text for details. \label{Tab:RW2d}}
\end{table}
Although large fluctuations are still present, we may notice
that both PENTA89 and ZIFF89 exhibit correlated results with
$n_c \approx 95 - 200$. In order to quantify this result
more precisely, we devised the $n$-block test, whose results
we will consider next. Because the results of the random walk
test and the $n$-block test are very closely related to each
other, discussion of the random walk test will be given
together with the $n$-block test at the end of the
following Section.

\bigskip
\section{$n$-block test}

In the $n$-block test, we used an approach similar to the random
walk test. First, we studied various generators with
parameters $n=10^4$ and $N=10^6$. In the cases of GGL, RAND, RANMAR
and RAN3, we observed no correlations. Then, for GFSR generators
R31, R250 and R521 we performed an iterative study by varying $n$.
When $N=10^6$ samples were taken, the resulting correlation lengths
$n_c$ were $32 \pm 1$, $267 \pm 5$, and $555 \pm 5$, respectively.
With better statistics $N=10^8$, we observed no change for R31,
whereas the estimate for R521 reduced to $525 \pm 1$, and that of
R250 to $251 \pm 1$. The latter value was confirmed with $N=10^9$
also. Typical values of $\chi^2$ for R250 are shown in Fig.
\ref{Fig:nblock_r250}, where a sharp onset of correlations at
$n_c$ is visible.
\begin{figure}[htbp]
	\vspace{5.0cm}
\caption[The $\chi^2$ values for R250 in the $n$-block test.]
	{The $\chi^2$ values for R250 in the $n$-block test.
	Curves with circles and squares correspond
	to $N=10^8$ and $N=10^9$ samples, respectively. In both
	cases three independent runs have been performed.
	The horizontal line denotes $\chi^2=3.841$.
	\label{Fig:nblock_r250}}
\end{figure}

Then \hfill we concentrated \hfill on \hfill studying \hfill a
\hfill recommended \hfill generator \hfill ZIFF9689 \\
(GFSR(9689,471,314,157,$\oplus$)) \cite{Zif92,Zif92_pre}, which unlike
several other generators has performed well in recent simulations
of self-avoiding random walks \cite{Gra93a,Gra93b}.
This generator was extensively tested up to $n=25\,000$ and $N=10^7$,
but no correlations were found. In order to increase the number of
samples $N$, we tested ZIFF1279 (GFSR(1279,598,299,216,$\oplus$))
which is a 5-decimation of GFSR(1279,216,$\oplus$). With parameters
up to $n=1500$ and $N=10^9$, no correlations were observed.
These results suggest that deviations from random
behavior are less significant for ZIFF generators than for
GFSR generators with two lags. For
quantitative purposes, we have studied this subject in more
detail by comparing the results of R89, PENTA89, and ZIFF89.
These results are shown in Fig. \ref{Fig:RPZ_89}.
\begin{figure}[htbp]
\vspace{5.0cm}
\caption[The $\chi^2$ values for R89, PENTA89, and ZIFF89 as a
	function of block size $n$ in the $n$-block test.]
	{On the left are shown the $\chi^2$ values of R89, PENTA89, and
	ZIFF89 as a function of block size $n$, when $N=10^8$ samples
	have been taken. In the case of R89 results with $N=10^6$ samples
	(open triangles) are also shown. On the right results of
	PENTA89 and ZIFF89 have been compared with better
	statistics $N=10^9$. \label{Fig:RPZ_89}}
\end{figure}
The figure on the left clearly shows how dramatically
inferior GFSR generators based on primitive trinomials are
compared to generators which are based on either decimation
of such sequences or use of primitive pentanomials.
Furthermore, when PENTA89 and ZIFF89 are compared to each
other with higher statistics $N=10^9$ (figure on the right),
we may notice that at least in this particular case the
decimated sequence ZIFF89 performs somewhat better than
PENTA89, although correlations in both sequences are
clearly present.

The results of the random walk and $n$-block tests together
show that for the GFSR generators with two lags, the origin
of the errors in the simulations presented here and in Refs.
\cite{Cod93,Fer92,Gra93a,Gra93b,Sel93} must be the appearance
of local correlations in the probability distribution. Moreover,
although some empirical estimates for the correlation length have
previously been given \cite{Gra93a,Gra93b}, our tests are the first
which quantitatively show that the correlation length
lies very close to the expected value \cite{Com87,Zif92_pre}
of the longer lag parameter $p$. Furthermore, for the
generators based on a judicious decimation ({\em e.g.}
$k=3, 5, 7$) of GFSR generators (with two lags) or when
primitive pentanomials are used as a basis for a generator,
our results show that similar but less clear behavior is
observed for these generators as well. Thus, generators using
three consequtive exclusive-or operations seem to shuffle bits
better than R{\em p} generators in which only one exclusive-or
operation is used. This results from the fact that compared
with R{\em p} generators, in ZIFF{\em p} generators the
three-point correlations are much farther apart,
and therefore higher order correlations dominate their
behavior \cite{Zif92_pre}. Although we are not aware of
any theoretical studies for PENTA{\em p} generators,
we {\em assume} that this is what happens for them also.
In other words, the approach of using multiple
exclusive-or's does not remove the correlations but makes
them more subtle, as was also observed with the
autocorrelation test where both ZIFF89 and PENTA89
passed the test with $N=10^8$ samples, whereas R89 did not.
Therefore, we may conclude this part of our work by saying
that when generators based on the exclusive-or operation with
a low lag parameter $p$ are used, their results should be
taken with a very sceptical attitude. On the other hand, if
GFSR generators are still used, dubious results may not appear
if multiple exclusive-or's and greater lags
such as $p \geq 1279$ are used.

These results convincingly show that both the random walk test
and the $n$-block test are very well suited for detecting local
deviations from randomness in pseudorandom number sequences.
The $n$-block test in particular is very efficient due to
its simple nature. Moreover, it is important to realize
that many other applications such as studies of self-avoiding
random walks \cite{Sok94}, percolation phenomena \cite{Sta85},
and diffusion limited aggregation \cite{Vic89} are based on
the use of random walks. Therefore, results of our random walk
tests as well as the autocorrelation test are valid for such
applications as well.

\bigskip
\section{Condition number test}

The studies of the condition number test consist of two parts.
First, we will study the quality for three generation methods of
Gaussian distributed random variables. Then, the best method
will be used in further studies where several pseudorandom
number generators will be tested using the condition number test.

\bigskip
\subsection{Testing generation methods for Gaussian distributed
            random variables}

The first part of this test consists of testing the quality of
three generation methods of Gaussian distributed random variables:
Marsaglia's polar method, the ratio method, and a method
based on the central limit theorem. We first used parameters
$m=10$, $N=10\,000$, and $M=100$. With this choice of parameters,
three independent tests for each method were carried out. In the
case of Marsaglia's polar method the results for $\epsilon$ were
0.01, 0.005, and 0.02. In the case of the ratio method we found
values 0.015, 0.015, and 0.03. Therefore, no significant deviations
from the expected behavior with these methods were found. In the
case of the method based on the central limit theorem, however, the
number $n$ of U(0,1) distributed random numbers (used in the
generation of one Gaussian distributed random variable) affected
the quality of the Gaussian distribution significantly.
This is illustrated in Fig. \ref{Fig:CentralLT} where
$\epsilon$ is shown as a function of $n$.
\begin{figure}[htbp]
	\vspace{5.0cm}
\caption[Some results of the condition number test for the method
	based on the central limit theorem.]
	{The test statistic $\epsilon$ as a function of $n$
	used uniform random numbers, when Gaussian distributed
	random variables have been generated with the method
	based on the central limit theorem. ``Good'' values of
	$\epsilon$ are $0.02 \pm 0.01$. The inner figure
	contains a small portion of the main figure.
	\label{Fig:CentralLT}}
\end{figure}
Clearly the often suggested value $n=12$ is not large
enough for a good approximation of the Gaussian distribution.
Instead of that, values of $n \geq 96$ should be used.

Additional tests with parameters $m=10$, $N=10\,000$,
and $M=1000$ were also performed. The results for
Marsaglia's polar method were 0.0105, 0.0205, and 0.02,
while the ratio method gave 0.0245, 0.0235, and 0.023.
Since Marsaglia's polar method is faster than the
ratio method, the former was chosen for further studies
of the condition number test in this work.

\bigskip
\subsection{Testing generators}

In the condition number test, generators R31, R250, GGL, RAND,
RAN3, and RANMAR were tested with two sets of parameters. First,
we used $m=10$, $N=1000$, and $M=1000$. None of the generators
failed, when a single test was carried out. Then, with another
set of parameters $m=100$, $N=1000$, and $M=100$, generators
R31 and R250 passed the test, whereas GGL, RAND, RAN3, and
RANMAR failed the test. When the same test was repeated with a
different initial seed value, all the remaining generators
passed the test also.

Based on the results of the condition number test, subtle
deviations from randomness in the output of uniform
pseudorandom number generators are not very significant
when such random numbers are being transformed into Gaussian
distributed variables. One may assume that due to calculation
of several arithmetic operations and functions, the transformation
method reduces the effects of inevitable correlations, and if
such a method is good enough, even relatively poor uniform
random number generators may do well in applications in which
Gaussian distributed random variables are generated.


\chapter{Summary and discussion}

In addition to traditional applications such as lottery
and numerical integration, random numbers are needed
in numerous modern applications: high precision Monte
Carlo simulations in physical sciences \cite{Bin92},
image compression \cite{Barnsley88}, algorithm development
\cite{Ede_comm}, and stochastic optimization \cite{Aar89},
to name but a few. All these methods need
reliable but still practical sources of random numbers,
which due to practical reasons are usually generated
by means of deterministic algorithms, implemented as
pseudorandom number generators. Since
the reliability of the results in all these applications
depends on the quality of random numbers, there must be
efficient means to confirm that. These means are empirical
and theoretical tests. Although theoretical
tests based on studying some general properties of
pseudorandom number sequences give us basic knowledge of
the properties of such sequences, random number testing
is still mainly an empirical science. Though no empirical test
can ever {\em prove} ``goodness'' of the quality of random number
sequences, such tests give us a valuable insight into their
properties. However, since the number of possible tests is
practically unlimited, it is important that many tests,
which mimic the properties of the application in which the
random number sequences will be used, are included in the
test program. This idea leads to the concept of application
specific testing.

The motivation of this work has been twofold. First, we
have wanted to develop efficient tests for random numbers,
which are used in high precision physical simulations.
This need arises from the fact that presently there
are only few efficient empirical tests which mimic
the properties of commonly studied simulation models.
Second, recent high precision Monte Carlo simulations
\cite{Cod93,Fer92,Gra93a,Gra93b,Sel93}
have revealed anomalous correlations in the output of
some commonly used pseudorandom number generators, and
in that of the so called generalized feedback shift
register (GFSR) generators in particular. Although there
is good reason to assume that the erroneous results of
GFSR generators are due to the three-point correlations
with a correlation length $\xi = p$ (cf. Section 3.1),
efficient test methods for confirming this have not
been available up to date.

In this work, we have presented five simple tests
for detecting correlations in random number sequences.
The cluster test is based on the idea of comparing the
cluster size distribution of a random lattice with the
Ising model at an infinite temperature.
Another test based on the use of the Ising model is the
autocorrelation test, in which integrated autocorrelation
times of some quantities of the Ising model are calculated.
Then, in the random walk test we follow a random walker for
a certain number of steps on a plane. The $n$-block test is
also based on random walks, although in a simpler way.
Finally, the condition number test uses some properties
of random matrices, which are widely used in several
applications such as nuclear physics \cite{Mehta67}
and studies of chaotic systems \cite{Crisanti93}.

We implemented the cluster test on bit level and found it
to be very effective in localization of periodic correlations
in individual bits of random number sequences. It is remarkable,
however, that correlations in GFSR generators were not observed,
which shows that these generators should be good enough for
many applications in which good bit level properties
{\it of their individual bits} are required. However, since
crosscorrelations between various bits in random numbers
were not studied in this work, we do not want to rule out a
possibility of bit level correlations either. Other tests
were implemented to study random words. The results of the
autocorrelation test, random walk test, and $n$-block
test were especially enlightening, since these tests revealed
the reason for poor behavior of GFSR generators in Refs.
\cite{Cod93,Fer92,Sel93}.
First, by means of the autocorrelation test we
found that the origin of the errors observed in Ref. \cite{Fer92}
lies in local correlations present in the cluster formation
process of the Wolff algorithm \cite{Wol89a,Wol89b} since
the average cluster sizes in these cases were clearly biased.
Moreover, the integrated autocorrelation times of some
quantities in the Ising model were found very sensitive measures of
correlations. Then, based on the arguments in Refs.
\cite{Com87,Zif92_pre},
the correlation length of GFSR generators in the case of
three-point correlations equals their longer lag $p$, which might
be the real reason for the aforementioned errors.
In this work, by means of the random walk and $n$-block tests, we
have been able to reveal that this is indeed the case.
Therefore, the random walks and $n$-block
tests are very efficient in detecting
local correlations in random number sequences. Furthermore,
the problems observed in Refs. \cite{Fer92,Sel93} concern
mainly GFSR generators (with two lags) using one exclusive-or
operation, since similar lagged Fibonacci generators using
one of operations $\{+,-,\times\}$ have given correct results
in corresponding simulations \cite{Cod93}. One way to
avoid this problem is to use four lags instead of two,
as we have shown in this work. Such generators can be formed
by using tables of primitive pentanomials \cite{Kur91} or by
decimating GFSR sequences \cite{Zif92,Zif92_pre}; {\em i.e.}
taking every third number of their sequence, for example.
Such approaches do not eliminate the three-point correlations
but make them more subtle (farther apart), and therefore
improve the quality of generated random numbers.
Such generators have passed all our tests, when
the longest lag parameter $p$ has been chosen large
enough ($p \geq 1279$). Finally, we have also performed
tests for Gaussian distributed random matrices, but
our results show that most Gaussian transformation methods
reduce the effects of inevitable correlations in (uniformly
distributed) random numbers to such an extent, that with our
set of test parameters in the condition number test,
no deviations from randomness with any of the generators were
found. An exception was observed in the case of the transformation
method based on the central limit theorem. We found this method
to converge very slowly towards Gaussian distribution.

In conclusion, we believe that the tests for randomness presented
here form an efficient test bench which can be used to
develop better generators for demanding applications in
physical sciences. Our results of the decimated sequences
of GFSR generators are one step towards this direction.
However, we note that it is still of crucial importance
to further develop application specific tests along the
lines presented here to detect more subtle correlations,
which may not be revealed by the present test methods.

\appendix


\chapter{Polynomials $D_{s}(p)$}

\begin{table}
\vspace{-10.0cm}
\small
\begin{eqnarray*}
\left\{ \begin{array}{ccl}
       D_{1}(p) & = & p^4  \\
       D_{2}(p) & = & p^6  \\
       D_{3}(p) & = & p^9 + p^7  \\
       D_{4}(p) & = & p^{12} + p^{11} + p^9  + p^8 \\
       D_{5}(p) & = & p^{15} + p^{13} + p^{12} + p^{11} + p^8 \\
       D_{6}(p) & = & p^{18} + p^{15} + p^{14} + p^{13} + p^{11} + p^{10} +
          p^9  \\
       D_{7}(p) & = & p^{21} + p^{19} + p^{18} + p^{14} + p^{13} + p^{12} +
          p^{11} + p^9  \\
       D_{8}(p) & = & p^{24} + p^{21} + p^{19} + p^{18} + p^{16} + p^{13} +
          p^{12} + p^9  \\
       D_{9}(p) & = & p^{27} + p^{25} + p^{23} + p^{22} + p^{20} + p^{18} +
	              p^{17} + p^{16} + p^{15} + p^{14} + p^{13} + p^{12}
		      \\
                &   & + p^{11} + p^{10}  \\
       D_{10}(p) & = & p^{30} + p^{27} + p^{26} + p^{23} + p^{22} + p^{20} +
          p^{19} + p^{18} + p^{14} + p^{13} + p^{11} + p^{10}  \\
       D_{11}(p) & = & p^{33} + p^{31} + p^{30} + p^{29} + p^{28} + p^{24} +
	               p^{23} + p^{22} + p^{21} + p^{20} + p^{18} + p^{16}
		       \\
		 &   & + p^{14} + p^{13} + p^{12} + p^{10}  \\
       D_{12}(p) & = & p^{36} + p^{33} + p^{31} + p^{30} + p^{29} + p^{27} +
                       p^{25} + p^{24} + p^{22} + p^{21} + p^{20} + p^{18}
		       \\
                 &   & + p^{16} + p^{12} + p^{10}  \\
       D_{13}(p) & = & p^{39} + p^{37} + p^{33} + p^{32} + p^{29} + p^{27} +
                       p^{26} + p^{23} + p^{19} + p^{17} + p^{16} + p^{15}
		       \\
                 &   & + p^{14} + p^{10}  \\
       D_{14}(p) & = & p^{42} + p^{39} + p^{36} + p^{34} + p^{33} + p^{30} +
                       p^{29} + p^{27} + p^{26} + p^{25} + p^{23} + p^{22}
		       \\
                 &   & + p^{21} + p^{20} + p^{17} + p^{16} + p^{15} +
			 p^{14} + p^{13} + p^{12} + p^{11}  \\
       D_{15}(p) & = & p^{45} + p^{43} + p^{42} + p^{40} + p^{37} + p^{36} +
                       p^{34} + p^{31} + p^{26} + p^{25} + p^{23} + p^{22}
		       \\
                 &   & + p^{21} + p^{19} + p^{18} + p^{17} + p^{13} +
			 p^{12} + p^{11}  \\
       D_{16}(p) & = & p^{48} + p^{45} + p^{44} + p^{42} + p^{41} + p^{40} +
                       p^{39} + p^{38} + p^{29} + p^{28} + p^{27} + p^{26}
		       \\
	         &   & + p^{25} + p^{21} + p^{18} + p^{16} + p^{15} +
			 p^{12} + p^{11}  \\
       D_{17}(p) & = & p^{51} + p^{49} + p^{47} + p^{45} + p^{43} + p^{41} +
                       p^{39} + p^{38} + p^{37} + p^{36} + p^{33} + p^{32}
		       \\
                 &   & + p^{31} + p^{30} + p^{25} + p^{24} + p^{23} +
			 p^{22} + p^{20} + p^{19} + p^{18} + p^{15} +
			 p^{14} + p^{13}
		       \\
                 &   & + p^{11}  \\
\end{array} \right.
\end{eqnarray*}
\caption{The polynomials $D_{s}(p)$ in $p=1/2$ \protect\cite{Syk76}.}
\end{table}


\chapter{Results of the first cluster test}

\begin{table}[h]
\scriptsize \centering
\begin{tabular}{| r | c | c | c | c | c | c |}
\hline\hline
\multicolumn{1}{| c |}{ } &
\multicolumn{6}{ c |}{Random number generator} \\
\cline{2-7}
  \multicolumn{1}{| c |} {Bit}
& \multicolumn{1}{c |}{\makebox[1.5cm]{GGL}}
& \multicolumn{1}{c |}{\makebox[1.5cm]{R250}}
& \multicolumn{1}{c |}{\makebox[1.5cm]{R1279}}
& \multicolumn{1}{c |}{\makebox[1.5cm]{RANMAR}}
& \multicolumn{1}{c |}{\makebox[1.5cm]{RAN3}}
& \multicolumn{1}{c |}{\makebox[1.5cm]{RAND}} \\ \hline

 1 & 1.906 & 0.697 & 0.733 & 1.640 	& {\bf 18.39} 	& 0.853 \\
 2 & 0.215 & 0.447 & 0.185 & 0.131 	& {\bf 19.85} 	& 0.001 \\
 3 & 1.005 & 0.259 & 0.420 & 0.254 	& {\bf 20.26} 	& 0.804 \\
 4 & 2.160 & 0.815 & 0.186 & 0.649 	& {\bf 11.18} 	& 0.613 \\
 5 & 0.389 & 0.054 & 1.523 & 0.921 	& 0.602 	& 0.959 \\
 6 & 0.846 & 0.685 & 0.192 & 1.548 	& 0.679 	& 1.045 \\
 7 & 1.787 & 0.447 & 0.838 & 0.340 	& 0.599 	& {\bf 4.290} \\
 8 & 1.666 & 1.440 & 0.040 & 0.234 	& 0.341 	& {\bf 6.929} \\
 9 & 0.122 & 1.134 & 0.768 & 0.579 	& 0.082 	& {\bf 6.533} \\
10 & 1.284 & 1.221 & 0.864 & 0.494 	& 1.107 	& {\bf 7.187} \\
11 & 1.734 & 0.939 & 0.298 & 1.274 	& 0.488 	& {\bf 13.29} \\
12 & 0.437 & 0.173 & 0.457 & 0.580 	& 0.798 	& {\bf 4.424} \\
13 & 0.027 & 0.244 & 0.985 & 0.803 	& 0.207 	& {\bf 8.829} \\
14 & 2.557 & 0.558 & 1.032 & 0.635 	& 0.535 	& {\bf 89.09} \\
15 & 0.782 & 1.223 & 0.740 & {\bf 3.335} & 1.310 	& {\bf 142.2} \\
16 & 0.324 & 0.042 & 0.231 & 0.523 	& 0.911 	& {\bf 193.0} \\
17 & 1.428 & 1.572 & 0.767 & 0.131 	& 0.470 	& {\bf 128.0} \\
18 & 1.959 & 1.079 & 0.013 & 0.132 	& 1.570 	& {\bf 929.7} \\
19 & 0.199 & 0.075 & 1.097 & 0.883 	& 0.675 	& {\bf 1485.} \\
20 & 0.759 & 0.164 & 1.750 & 1.190 	& 1.276 	& {\bf 2812.} \\
21 & 2.266 & 0.809 & 0.613 & 1.098 	& 0.489 	& $\infty$ \\
22 & 0.977 & 2.134 & 0.433 & 0.076 	& 0.181 	& {\bf 5968.} \\
23 & 0.694 & 0.907 & 1.066 & 0.386 	& 1.520 	& {\bf 8868.} \\
24 & 1.663 & 1.034 & 1.309 & 0.396 	& 1.922 	& $\infty$ \\
25 & 1.945 & 0.247 & 1.540 & $\infty$ 	& {\bf 4424.} 	& $\infty$ \\
26 & 0.153 & 0.072 & 1.055 & $\infty$ 	& $\infty$	& $\infty$ \\
27 & 0.882 & 0.226 & 0.183 & $\infty$ 	& $\infty$  	& $\infty$ \\
28 & 2.283 & 1.694 & 1.161 & $\infty$ 	& $\infty$  	& $\infty$ \\
29 & 0.416 & 1.325 & 0.303 & $\infty$ 	& $\infty$  	& $\infty$ \\
30 & 0.567 & 0.675 & 0.087 & $\infty$ 	& $\infty$  	& $\infty$ \\
31 & 1.382 & 1.600 & 1.726 & $\infty$ 	& $\infty$  	& $\infty$ \\
\hline\hline
\end{tabular}
\caption[The values of goodness $g_i'$, $i=1,2,\ldots,31$, for the first
	run of the cluster test.]
	{The values of goodness $g_i'$, $i=1,2,\ldots,31$, for the first
	run of the cluster test. $\infty$ denotes a very large number
	(greater than $10^5$). Other failing results are given in bold
	type. See text for details.}
\end{table}


\chapter{Results of the second cluster test}

\begin{table}[h]
\scriptsize \centering
\begin{tabular}{| r | c | c | c | c | c | c |}
\hline\hline
\multicolumn{1}{| c |}{ } &
\multicolumn{6}{ c |}{Random number generator} \\
\cline{2-7}
  \multicolumn{1}{| c |} {Bit}
& \multicolumn{1}{c |}{\makebox[1.5cm]{GGL}}
& \multicolumn{1}{c |}{\makebox[1.5cm]{R250}}
& \multicolumn{1}{c |}{\makebox[1.5cm]{R1279}}
& \multicolumn{1}{c |}{\makebox[1.5cm]{RANMAR}}
& \multicolumn{1}{c |}{\makebox[1.5cm]{RAN3}}
& \multicolumn{1}{c |}{\makebox[1.5cm]{RAND}} \\
\hline
 1 & 1.806 & 1.218 & 0.666 & 1.355 & {\bf 21.03}	& 0.606 \\
 2 & 1.888 & 0.383 & 0.535 & 0.219 & {\bf 20.61}	& 0.364 \\
 3 & 0.117 & 0.496 & 1.678 & 1.584 & {\bf 20.15}	& 0.656 \\
 4 & 1.188 & 0.038 & 0.715 & 1.498 & {\bf 14.62}	& 0.564 \\
 5 & 2.364 & 1.627 & 1.885 & 1.995 	& 0.130 	& 0.980 \\
 6 & 0.710 & 0.059 & 1.874 & 0.073 	& 1.338 	& 1.682 \\
 7 & 0.291 & 0.494 & 1.829 & 1.667 	& 0.038 	& {\bf 4.282} \\
 8 & 1.338 & 1.275 & 1.542 & 1.233 	& 0.121 	& {\bf 7.253} \\
 9 & 1.171 & 1.936 & 0.391 & 0.916 	& 0.807 	& {\bf 6.374} \\
10 & 0.020 & 1.384 & 1.079 & 1.090 	& 2.280 	& {\bf 7.690} \\
11 & 1.227 & 0.055 & 1.864 & 0.359 	& 1.356 	& {\bf 12.40} \\
12 & 1.876 & 0.205 & 0.461 & 1.948 	& 0.767 	& {\bf 4.348} \\
13 & 0.438 & 1.471 & 1.277 & 0.132 	& 2.492 	& {\bf 9.594} \\
14 & 0.225 & 0.290 & 0.743 & 1.202 	& 0.039 	& {\bf 93.68} \\
15 & 2.408 & 0.018 & 0.406 & 0.546 	& 1.866 	& {\bf 144.8} \\
16 & 0.824 & 0.130 & 0.903 & 1.227 	& 0.766 	& {\bf 209.6} \\
17 & 0.456 & 0.472 & 1.602 & 0.812 	& 1.429 	& {\bf 141.5} \\
18 & 1.922 & 0.008 & 1.873 & 1.548 	& 0.161 	& {\bf 1063.} \\
19 & 2.313 & 1.489 & 0.460 & 0.113 	& 0.681 	& {\bf 1343.} \\
20 & 0.288 & 0.362 & 1.835 & 0.764 	& 0.111 	& {\bf 2858.} \\
21 & 1.304 & 0.056 & 0.612 & 0.789 	& 1.069 	& $\infty$ \\
22 & 2.598 & 0.646 & 2.704 & 0.467 	& 1.268 	& {\bf 6458.} \\
23 & 1.412 & 1.370 & 0.816 & 0.798 	& 0.285 	& {\bf 5720.} \\
24 & 0.851 & 0.535 & 1.544 & 0.755 	& 1.151 	& $\infty$ \\
25 & 1.712 & 0.536 & 0.797 & $\infty$	& {\bf 4542.}	&  $\infty$  \\
26 & 2.198 & 0.024 & 0.172 & $\infty$	& $\infty$  	&  $\infty$  \\
27 & 0.044 & 1.372 & 0.993 & $\infty$	& $\infty$  	&  $\infty$  \\
28 & 1.124 & 0.163 & 0.544 & $\infty$	& $\infty$  	&  $\infty$  \\
29 & 1.995 & 0.179 & 0.185 & $\infty$	& $\infty$  	&  $\infty$  \\
30 & 0.029 & 0.250 & 1.389 & $\infty$	& $\infty$  	&  $\infty$  \\
31 & 0.531 & 0.692 & 1.230 & $\infty$	& $\infty$  	&  $\infty$  \\
\hline\hline
\end{tabular}
\caption[The values of goodness $g_i'$, $i=1,2,\ldots,31$, for the second
	run of the cluster test.]
	{The values of goodness $g_i'$, $i=1,2,\ldots,31$, for the second
	run of the cluster test. $\infty$ denotes a very large number
	(greater than $10^5$). Other failing results are given in bold type.
	See text for details.}
\end{table}


\begin{thebibliography}{250}
\bibitem{Ric92}
{\sc M.\ Richter} has generated millions of physical random
numbers (known as PURAN II) by measuring thermal noise of a
semiconductor device. Statistical properties of these numbers
were tested in Ref. \cite{Vat93}, where no errors were found.
The random numbers by PURAN II are available from an anonymous
FTP site at {\tt dfv.rwth-aachen.de}.
\bibitem{Sta94}
{\sc D. Stauffer} gave us the main part of the Ziff
pseudorandom number generator \cite{Zif92},
which we extended to take care of initialization and
vectorization; personal communication.
\end{thebibliography}
\end{document}